\documentclass[preprint]{aastex}

\usepackage[dvips]{color}
\usepackage{here}
\usepackage{amsmath,amssymb}
\slugcomment{}
\shortauthors{Xue \& Suto}
\shorttitle{Coplanar Hierarchical Triple Systems}
\begin{document}
\title{Difficulty in Formation of Counter-orbiting Hot Jupiters from
Near-coplanar Hierarchical Triple Systems: A Sub-stellar Perturber}
\author{
Yuxin Xue\altaffilmark{1} and Yasushi Suto\altaffilmark{1,2}
} 
\affil{${}^1$Department of Physics, 
The University of Tokyo, Tokyo 113-0033, Japan \\
${}^2$ Research Center for the Early Universe, School of Science, 
The University of Tokyo, \\ Tokyo 113-0033, Japan}

\email{yuxin@utap.phys.s.u-tokyo.ac.jp}

\begin{abstract}

Among a hundred transiting planets with a measured projected spin-orbit
angle $\lambda$, several systems are suggested to be
counter-orbiting. While they may be due to the projection effect, the
mechanism to produce a counter-orbiting planet is not established.  A
promising scenario for the counter-orbiting planets is the extreme
eccentricity evolution in near-coplanar hierarchical triple systems with
eccentric inner and outer orbits. We examine this scenario in detail by
performing a series of systematic numerical simulations, and consider
the possibility of forming hot Jupiters, especially counter-orbiting one
under this mechanism with a distant sub-stellar perturber. We
incorporate quadrupole and octupole secular gravitational interaction
between the two orbits, and also short-range forces (correction for
general relativity, star and inner planetary tide and rotational
distortion) simultaneously. We find that most of systems are tidally
disrupted and that a small fraction of survived planets turns out to be
prograde.  The formation of counter-orbiting hot Jupiters in this
scenario is possible only in a very restricted parameter region, and
thus very unlikely in practice.
\end{abstract}

\clearpage

\section{Introduction \label{sec:intro}}

Ever since the first discovery of an exoplanet, 51 Peg b, more than 100
Hot Jupiters (HJs) with semi-major axis $<0.1$ AU have been detected
around main-sequence stars. Nevertheless their origin remains as one of
the important unsolved puzzles in this field.  It is generally
believed that such gas giants are unlikely to be formed in-situ, and
instead, that they are first formed at large distance from the central
star, most likely beyond the ice line, and then migrated significantly
inward to the current orbits (but see, e.g.,
\citet{Boley2015} and \citet{Batygin2015} for different ideas).

The migration mechanisms are not yet established, but possible
scenarios include (1) disk-planet interaction \citep[e.g.,][]{Lin1996,
Alibert2005}, (2) planet-planet scattering \citep[e.g.,][]{Rasio1996,
Nagasawa2008, Nagasawa2011, Beauge2012}, (3) the Lidov-Kozai
migration \citep[e.g.,][]{Lidov1962, Kozai1962, Wu2003, Fabrycky2007,
Naoz2011, Petrovich2015a, Anderson2016}, and (4) secular migration
\citep{Wu2011}.

In reality, those different migration mechanisms may have contributed to
the observed HJ population to some degree. Each mechanism often predicts
different statistical distribution and correlations of the resulting
orbital parameters of the planetary systems, and the relevant
observations may provide a potential clue to distinguish different
mechanisms. For example, disk-planet interaction would imply that gas
giants smoothly migrate inward in a gaseous disk and thus the angle,
$\psi$, between the stellar spin and planetary orbital axes would not
significantly change from its initial value (most likely very close to
zero, but it is possible that the spin axis of the central
star is moderately misaligned with the normal vector of the primordial
disk \citep[e.g.,][]{Bate2010, Foucart2011, Batygin2012, Lai2014}. In contrast, the other migration mechanisms
mentioned above rely on a dynamical process after the depletion of the
gas disk, which can induce a strong spin-orbit misalignment.  For this
reason, measurement of $\psi$ can be a useful probe in understanding the
origin of HJs.

Indeed, the Rossiter-McLaughlin effect has been very successful in
measuring the sky projected spin-orbit angle, $\lambda$ for transiting
planetary systems \citep{Rossiter1924,McLaughlin1924,Queloz2000,Ohta2005,Winn2005}.
Approximately one-third of the measured systems exhibit significant
misalignment of $\lambda>\pi/4$, and a dozen of systems turned out to be
even in a retrograde orbit ($\lambda> \pi/2$); see Fig.7 of
\citet{Xue2014} for example. Such unexpected and counter-intuitive
discoveries imply that those HJs should have experienced violent
dynamical processes.

While all the above three dynamical migration mechanisms could produce
retrograde HJs, it has been shown difficult to produce counter-orbiting
HJs \citep[e.g,][]{Fabrycky2007,Naoz2011,Liu2015,Petrovich2015b}. For
definiteness, we call the {\it counter-orbiting} planets as those with
$160^{\circ}<\psi<180^{\circ}$, and {\it retrograde} planets are simply
used to indicate $\psi>\pi/2$ throughout the present paper even if the
distinction may not be conventional.

In this context, we should note that the observed $\lambda$ differs from
the {\it true} spin--orbit angle $\psi$; they are related in terms of
the orbital inclination $i_{\rm orb}$ and the obliquity of the stellar
spin-axis $i_\star$ as
\begin{equation}
\label{eq:psi-lambda}
\cos\psi = \cos i_\star \cos i_{\rm orb} 
+ \sin i_\star \sin i_{\rm orb} \cos\lambda 
\approx \sin i_\star \cos\lambda .
\end{equation}
The above approximation holds for transiting systems with $i_{\rm obs}
\approx \pi/2$. Since the stellar axis is usually defined so that
$0<i_\star<\pi/2$, equation (\ref{eq:psi-lambda}) implies that $\psi \ge
\lambda$ if $0<\lambda<\pi/2$ while $\psi \le \lambda$ if $\pi/2
<\lambda<\pi$.

The true spin-orbit angle $\psi$ is not so easy to obtain, but can be
estimated by combining the measurement of $i_\star$ via asteroseismology
\citep{Unno1989,Gizon2003,Huber2013,Campante2014,Christensen-Dalsgaard2014}.
\citet{Benomar2014} performed the first quantitative determination of
$\psi$ for transiting planetary systems around main-sequence stars. For
HAT-P-7, their asteroseismology analysis yields $i_\star\approx
30^\circ$, and they obtain $\psi \approx 120^\circ$ from the joint
analysis of the Rossiter-McLaughlin measurement of $\lambda \approx
180^\circ$.  For Kepler-25c, they obtain
$i_\star=65^\circ.4_{-6^\circ.4}^{+10^\circ.6}$, and $\psi =
26^\circ.9_{-9^\circ.2}^{+7^\circ.0}$, which should be compared with
$\lambda = 9^\circ.4 \pm {7^\circ.1}$.  Indeed these results demonstrate
the importance of the projection effect mentioned above. More
importantly, planetary systems with $\lambda \approx 180^\circ$ may not
be necessarily counter-orbiting, but just retrograde.  This may also be
the case for HAT-P-6b with $\lambda = 165^\circ \pm
{6^\circ}$\citep{Albrecht2012}, HAT-P-14b with $\lambda = 189^{\circ}.1
\pm {5^\circ.1}$\citep{Winn2011}.

Therefore the existence of the counter-orbiting planets has not yet been
established observationally so far. Nevertheless, it is tempting to
consider a dynamical model that can theoretically explain the
counter-orbiting HJs if exist at all.  One promising mechanism has been
recently proposed by \citet{Li2014}.  They consider a near-coplanar
hierarchical triple system, and derived a {\it flip condition} that
the inner planet becomes counter-orbiting under the secular perturbation
up to the octupole-order of the gravitational potential of the outer
object in a very eccentric orbit. 

To be more specific, their flip condition is written as
\begin{eqnarray} 
\label{eq:flipcondition} 
\epsilon > \epsilon_{\rm crit,i}
= \frac{8}{5} 
\frac{1-e_{1,i}^{2}}{7-e_{1,i}(4+3e_{1,i}^{2})\cos(\omega_{1,i}+\Omega_{1,i})},
\end{eqnarray}
in terms of 
\begin{eqnarray} 
\label{eq:epsilondefinition} 
\epsilon \equiv \frac{a_{1}}{a_{2}}\frac{e_{2}}{1-e_{2}^{2}}
\end{eqnarray}
that characterizes the ratio of the orbit-averaged octupole to
quadrupole terms in the massless limit ($m_{1} \ll m_{0}, m_{1} \ll
m_{2}$). In the above expressions, $e$, $a$, $\omega$, $\Omega$, and $m$
denote the eccentricity, semi-major axis, argument of periastron,
longitude of ascending node with the subscripts 1 and 2 indicating the
inner and outer body, respectively. In the massless limit, $a_1$, $a_2$,
and $e_2$ are conserved, and thus $\epsilon$ defined by equation
(\ref{eq:epsilondefinition}) is also a constant of motion. The other
orbital elements are time-dependent, and we use the subscript $i$ in
equation (\ref{eq:flipcondition}) in order to indicate their initial
values.

\cite{Petrovich2015b} presented a more general form of the flip condition
(\ref{eq:flipcondition}) on the basis of the conservation of the energy
(the orbit-averaged quadrupole and octupole potential terms) for the
coplanar hierarchical triple configuration. His result, equation (11) of
\cite{Petrovich2015b}, can be written as
\begin{eqnarray} 
\label{eq:epsilon-oct} 
\epsilon_{\rm L} < \epsilon_{\rm oct} \equiv 
\frac{m_0-m_1}{m_0+m_1}\frac{a_{1}}{a_{2}}\frac{e_{2}}{1-e_{2}^{2}}
<\epsilon_{\rm U},
\end{eqnarray}
which reduces to equation (\ref{eq:flipcondition}) in the massless (or
test-particle) limit of the inner planet. The lower and upper
limits, $\epsilon_{\rm L}$ and $\epsilon_{\rm U}$, defining the boundary
of the flip region are determined by the value of the final angle
$\varpi_f \equiv \cos^{-1} \hat{\mathbf e}_{1,f} \cdot \hat{\mathbf e}_{2,f}$
between the inner and outer orbital unit Lenz vectors. Specifically 
$\epsilon_{\rm L}$ and $\epsilon_{\rm U}$ correspond to
$\varpi_f = 0$ and $\pi$, respectively. The upper limit can be
practically neglected for sub-stellar perturbers as considered in the
present study, but is very important for planetary perturbers
(Xue, Masuda, \& Suto, in preparation).

\citet{Li2014} numerically computed the evolution of such coplanar
triple systems in the massless limit, and confirmed that the flip
condition is very well described by equation (\ref{eq:flipcondition}).
Also, in the large inclination regime, that analytical flip criterion
agrees well with the numerical results even up to $m_{2}/m_{1} > 7$
\citep{Teyssandier2013}.  They found that $e_1$ increases monotonically
and the mutual orbital inclination between the two bodies, $i_{12}$,
remains low just before the flip, and then the orbital flip of the inner
planet proceeds in a very short timescale when $e_1$ becomes very close
to unity, $1-e_{1}\sim 10^{-3}-10^{-4}$.  In that case, the angular
momentum of the inner planet is roughly given as
$m_1\sqrt{Gm_0(1-e_1^2)} \approx m_1\sqrt{2Gm_0(1-e_1)}$, and even a
small perturbative torque may easily change the angular momentum of the
inner planet, and thus flip its orbit if the value of $1-e_1$ is
sufficiently small.  \citet{Li2014} proposed a coplanar-flip mechanism
for the formation of counter-orbiting HJs in which the inner planet
flips by $\sim 180^{\circ}$ before the tidal evolution dominates, and
then its extremely eccentric orbit is quickly circularized due to the
strong tidal interaction by the central star.

\citet{Liu2015} found, however, that the short-range forces,  General
Relativity(GR), planetary tide (non-dissipative) and rotational distortion suppress the
extreme value of $e_1$ that otherwise could be achieved due to the
octupole term in hierarchical triple systems with large mutual orbital
inclination (i.e., not coplanar configuration) in the
Lidov-Kozai oscillation.  These additional forces induce a precession
of the Lenz vector of the inner planet, and impose a strict upper limit
on the maximum achievable value of $e_1$; as the short-range forces
become stronger, the orbital flips are more confined to the region where
the mutual orbital inclination $i_{12}$ is close to $90^{\circ}$. This
result strongly implies that one needs to incorporate those short-range
forces in order to describe properly the dynamics of near-coplanar
hierarchical triple systems, which is not taken into account in
\citet{Li2014}.

\citet{Petrovich2015b} performed a series of such simulations for
planetary perturbers including short-range force effects that induce
pericenter precession of the inner orbit, such as GR, planetary tide and rotational distortion. All the
resulting HJs in his simulations turn out to be in a prograde and low
obliquity orbit. This is mainly because most of his initial conditions
do not satisfy $\epsilon_{\rm oct}<\epsilon_{\rm U}$ in the flip
condition (\ref{eq:epsilon-oct}) even when they satisfy $\epsilon_{\rm
oct}>\epsilon_{\rm L}$. Therefore his set of simulations does not cover
the relevant parameter space for the formation of counter-orbiting HJs
even though his simulations are for planetary perturbers, unlike for
sub-stellar perturbers as we consider below.

Those interesting previous results motivated us to systematically
explore the fate of the inner planet in near-coplanar hierarchical
triple systems including quadrupole and octupole terms of the
gravitational potential of the outer perturber, and short-range forces.
Our simulation is based on the orbit-averaged secular dynamics following
the formulation of \citet{Correia2011}, in which the stellar and
planetary spin effects are incorporated as well and the octupole order
effect is included following \citet{Liu2015}. Because we are primarily
interested in the origin of counter-orbiting HJs, we consider only those
systems that initially satisfy the analytical flip condition
(\ref{eq:flipcondition}). The present paper focuses on the stellar
perturber, and the parameter space relevant to the planetary perturber
will be discussed in the next paper. In this case, we find that most of
the systems are tidally disrupted and a fraction of survived planets
remains mainly as prograde HJs; the formation of counter-orbiting HJs is
possible only in a very restricted parameter range.

The rest of the paper is organized as follows. Section
\ref{sec:simulations} describes the basic configuration of the
hierarchical coplanar triple systems that we simulate. The simulation
results are presented and discussed in \S \ref{sec:dependence} where we
consider the parameter dependence in detail.  Section
\ref{sec:spin-orbit} presents the spin-orbit angle distribution in this
scenario.  Finally section \ref{sec:summary} is devoted to summary and
implications of the present paper. The set of equations that we employ
is based on \citet{Correia2011} and \citet{Liu2015}, but summarized
explicitly in \ref{sec:correia-eqs} for convenience and
definiteness. The analytical expression for the short-range force
effects are summarized in the \ref{sec:srf-eqs}.

\section{Numerical Simulations \label{sec:simulations}}

\begin{figure}[h]
\begin{center}
\includegraphics[width=15cm]{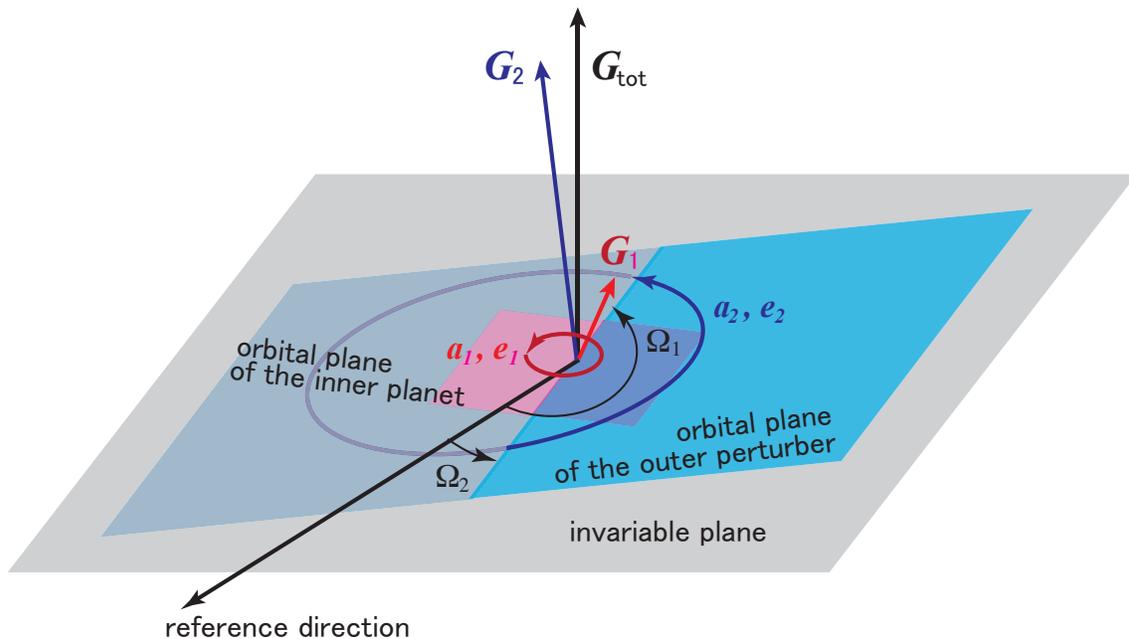} 
\caption{Schematic configuration of a near-coplanar triple system in the
Jacobi coordinate.}  \label{fig:schematic-triple}
\end{center}
\end{figure}

An schematic configuration of near-coplanar hierarchical triple systems
for our numerical simulations is illustrated in Figure
\ref{fig:schematic-triple}. A central star of mass $m_0$ and radius
$R_0$ is located at the origin of the coordinate.  The normal vector of
the invariable plane is the total orbital angular vector ${\mathbf
G}_{\rm tot}={\mathbf G}_{1}+{\mathbf G}_{2}$ of the inner and outer
bodies. Thus the mutual orbital inclination angle of the two orbits is
given by $i_{12}=i_1+i_2$ where $i_1$ and $i_2$ are the inclinations of
each orbit with respect to the invariable plane.  Throughout the present
paper, we adopt $m_{0} = 1M_{\odot}$, $R_{0} = 1R_{\odot}$, $m_{1} =
1M_{\rm J}$ and $R_{1} = 1R_{\rm J}$ for definiteness.

The equations of motions that we adopt are based on \citet{Correia2011},
in which short-range forces, GR, spin rotation and tidal effects for
both star and inner planet are included in addition to the quadrupole
term of the orbit-average gravitation potential of the outer body.  We
modify their equations so as to incorporate the the octupole secular
terms following \citet{Liu2015}.  The full equations of motion are
described in \ref{sec:correia-eqs}.

\bigskip

\begin{table}[h]
\begin{center}
\begin{tabular}{c||c|c|c|c|c|c||c|c|c|c} 
model &$a_{2}$(AU)& $e_{2}$ & $m_{2}(M_\odot)$ &$i_{12}$& $t_{\rm v,p}$(yr) &$f$&PHJ&RHJ&NM&TD\\ 
\hline
fiducial&$500$&0.6& $0.03$&$6^{\circ}$ &0.03&2.7&9.0\%&0.4\%&1.8\%&88.7\%\\ 
\hline
m001&$500$&0.6& $0.01$ &$6^{\circ}$ & 0.03&2.7&21.0\%&2.6\%&3.1\%&73.2\%\\
m010&$500$&0.6& $0.1$ &$6^{\circ}$ & 0.03&2.7&4.6\%&0.1\%&1.1\%&94.2\%\\
m100&$500$&0.6& $1$ &$6^{\circ}$ &
		     0.03&2.7&1.3\%&0.0\%&0.2\%&98.5\%\\ 
\hline
a200&$200$&0.6& $0.03$ &$6^{\circ}$ & 0.03&2.7&8.4\%&0.0\%&1.8\%&89.8\%\\
a100&$100$&0.6& $0.03$ &$6^{\circ}$ & 0.03&2.7&7.7\%&0.0\%&1.1\%&91.2\%\\
a050&$50$&0.6& $0.03$ &$6^{\circ}$ &
		     0.03&2.7&10.6\%&0.0\%&0.0\%&89.4\%\\ 
\hline
e03&$500$&0.3& $0.03$ &$6^{\circ}$ & 0.03&2.7&2.1\%&0.2\%&1.4\%&96.4\%\\
e04&$500$&0.4& $0.03$ &$6^{\circ}$ & 0.03&2.7&3.7\%&0.2\%&0.7\%&95.4\%\\
e05&$500$&0.5& $0.03$ &$6^{\circ}$ & 0.03&2.7&6.0\%&0.4\%&1.3\%&92.3\%\\
e07&$500$&0.7& $0.03$ &$6^{\circ}$ & 0.03&2.7&13.5\%&0.5\%&3.1\%&82.9\%\\
e08&$500$&0.8& $0.03$ &$6^{\circ}$ &
		     0.03&2.7&21.0\%&0.6\%&0.6\%&72.9\%\\ 
\hline
i30&$500$&0.6& $0.03$ &$30^{\circ}$ & 0.03&2.7&2.9\%&0.4\%&1.0\%&95.7\%\\
i15&$500$&0.6& $0.03$ &$15^{\circ}$ & 0.03&2.7&4.6\%&0.7\%&1.5\%&93.1\%\\
i00&$500$&0.6& $0.03$ &$0^{\circ}$ &
		     0.03&2.7&13.4\%&0.0\%&0.7\%&85.9\%\\ 
\hline
t03000&$500$&0.6& $0.03$ &$6^{\circ}$ & 0.3&2.7&3.4\%&0.0\%&2.2\%&94.4\%\\
t00030&$500$&0.6& $0.03$ &$6^{\circ}$ & 0.003&2.7&55.3\%&25.3\%&1.8\%&17.6\%\\
t00003&$500$&0.6& $0.03$ &$6^{\circ}$ &
		     0.0003&2.7&63.1\%&35.0\%&1.8\%&0.1\%\\ 
\hline
f216&500&0.6& $0.03$ &$6^{\circ}$ & 0.03&2.16&53.5\%&32.5\%&1.8\%&11.9\%\\
f166&500&0.6& $0.03$ &$6^{\circ}$ & 0.03&1.66&60.6\%&37.6\%&1.8\%&0.0\%\\
f000&500&0.6& $0.03$ &$6^{\circ}$ & 0.03&0.0&60.6\%&37.6\%&1.8\%&0.0\%\\ 
\hline
\end{tabular}
\caption{Summary of parameters and fates of our simulation runs.  All
the models adopt $m_{0} = 1M_{\odot}$, $R_{0} = 1R_{\odot}$, $m_{1} =
1M_{\rm J}$, $R_{1} = 1R_{\rm J}$, $\omega_{1,i} = 0$, $\omega_{2,i} = 0$,
$\Omega_{1,i} = \pi$, $\Omega_{2,i} = 0$, and $i_{s1,i} = 0$. The final
states are divided into four categories: Prograde HJ (PHJ; $a<0.1$ AU,
$e<0.01$, and $i_{\rm 12}<\pi/2$), Retrograde HJ (RHJ; $a<0.1$ AU,
$e<0.01$, and $i_{\rm 12}>\pi/2$), Non-migrating planets (NM) and
Tidally disrupted planets (TD; $q<R_{\rm roche}$). We performed 1800
runs over the grids of $\epsilon_i$ -- $e_{1,i}$ plane for each
model. \label{tab:simparam}}
\end{center}
\end{table}

\subsection{Model parameters \label{subsec:model}}

In the present paper, we have in mind a sub-stellar object as the outer
perturber. Specifically we adopt $a_{2,i} =500$ AU, $m_2=0.03M_\odot$,
$e_{2,i}=0.6$, $i_{12,i}=6^\circ$, the viscous time scale for the inner
planet, $t_{\rm v,p}=0.03$yr, and $f=2.7$.  The choice of those values
for the fiducial parameters is admittedly rather arbitrary because it is
very difficult to estimate their joint probability for actual
near-coplanar hierarchical triples.  Therefore we consider a variety of
simulation models with fixed $m_2$, $a_{2,i}$, $e_{2,i}$, $i_{12,i}$, $t_{\rm v,p}$, 
and $f$ as listed in Table \ref{tab:simparam}, instead of
sampling those parameters from their assumed distribution function.
Therefore our purpose is not to produce a mock distribution of real
near-coplanar hierarchical triples, but to understand the parameter
dependence of their dynamical evolution in a systematic fashion.

In each model, we perform 1800 different runs by varying
$(e_{1,i},\epsilon_{i})$ systematically; $e_{1,i}$ is varied between
$0.6$ and $0.96$ with a constant interval of $0.02$, and $\epsilon_{i}$
is varied between $\epsilon_{\rm crit,i}$ and 0.15 with a constant
interval of 0.001. Thus the value of $a_{1,i}$ in each run is uniquely
computed from $\epsilon_{i}$ through equation (\ref{eq:flipcondition}).
We note that in all the models, both $a_2$ and $e_2$ are practically
constant, i.e., $a_2=a_{2,i}$, and $e_2=e_{2,i}$, although $a_1$ and
$e_1$ significantly change from their initial values in most cases.

We fix the initial spin periods of the central star and inner planet as
$25\rm day$ and $10\rm day$, the viscous time scale of the star $t_{\rm
v,s}$ as 50 yr, and the Love numbers for the star and inner planet as
0.028 and 0.5, respectively.  The dimensionless principal moment of
inertia $I/(MR^2)$ of the star and inner planet are set to 0.08 and
0.26, respectively.  We do not randomly choose the initial phase angles
so that $\epsilon_{\rm crit,i}$ is independent of them in our parameter
survey; we adopt $\omega_{1,i} = 0$, $\omega_{2,i} = 0$, $\Omega_{1,i} =
\pi$, and $\Omega_{2,i} = 0$.  Since planets are generally expected to
form within a protoplanetary disk that is perpendicular to the spin
angular vector of the central star, the initial stellar inclination with
respect to the orbit of the inner planet is set to $i_{s1,i}=0$.

Following \citet{Petrovich2015a}, we divide the fate of the simulated
systems into four different categories, and stop the run when it reaches
one of the following states:
\begin{description}
\item[(i) PHJ (prograde HJ)]:
$a_{1,f} < 0.1$ AU, $e_{1,f} < 0.01$ and $i _{12,f}<\pi/2$. 
\item[(ii) RHJ (retrograde HJ)]:
$a_{1,f} < 0.1$ AU, $e_{1,f} < 0.01$ and $i _{12,f}>\pi/2$.
\item[(iii) TD (tidally disrupted within the Roche limit of the central
star)]: The inner planet is tidally disrupted if its pericenter distance
$q_1\equiv a_1(1-e_1)$ is less than the Roche limit:
\begin{eqnarray} 
\label{eq:roche-limit} 
q_1 < R_{\rm roche} \equiv f(m_{0}/m_{1})^{1/3}
= 0.0126 \left(\frac{f}{2.7}\right){\rm AU}
\sim 2.71\left(\frac{f}{2.7}\right)R_{\odot}.
\end{eqnarray}
The appropriate value for the Roche limit is somewhat uncertain. Thus
while our fiducial value of $f$ is 2.7 \citep[e.g,][]{Guillochon2011},
we consider $f=2.16$ \citep{Faber2005} and $f=1.66$ \citep{Naoz2012} as
well. Note, however, that $f \approx 1$ corresponds to the radius of the
central star itself, and the planet infalls to the star for $f<1$.
\item[(iv) NM (non-migrating planet)]
If the inner planet does not experience a significant migration, and 
stays at an orbit with
$a_{1,f} \sim a_{1,i}$ until $t=10^{10}$yr.
\end{description}
In the near-coplanar hierarchical triple systems as considered here, all
the survived PHJs and RHJs turn out to be very well aligned ($i_{12}
<10^\circ$) and counter-orbiting HJs ($i_{12} \sim \pi$), respectively.

Table \ref{tab:simparam} summarizes the model parameters of our
simulations as well as the fraction of their final states.  We should
emphasize here that the fraction of the final states listed in Table
\ref{tab:simparam} is computed assuming the uniform distribution over
the surveyed region of $e_{1,i}-\epsilon_{i}$ plane. In reality, it is
unlikely that $e_{1,i}$ and $\epsilon_{i}$ (or equivalently $a_{1,i}$)
are distributed uniformly. Nevertheless this is inevitable because we do
not assume any model-dependent and very uncertain prior distribution
function for $e_{1,i}$ and $\epsilon_{i}$ in this paper. Therefore the
values of {\it fraction} referred to throughout the present paper needs
to be interpreted with caution, but still provide an important measure
of the fate of the systems.

\subsection{Fiducial case \label{subsec:fiducial}}

\begin{figure}[h]
\begin{center}
\includegraphics[width=15cm]{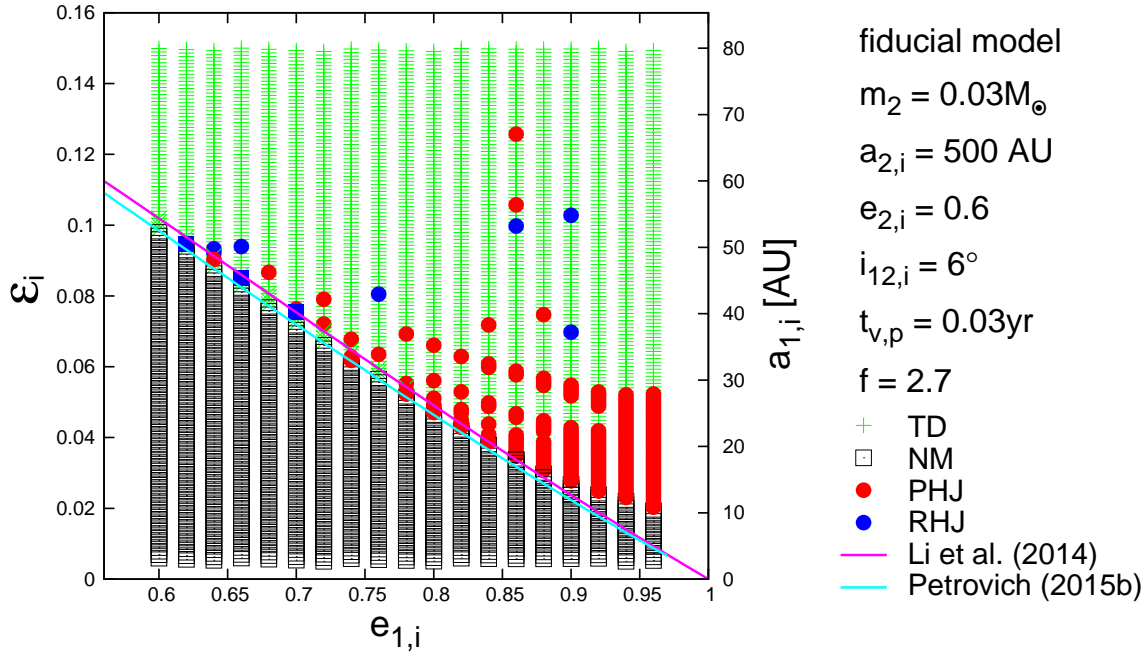} 
\caption{Fate of the inner planet on the $e_{1,i}-\epsilon_{1,i}$ plane
for our fiducial model; $a_{2,i} = 500$ AU, $m_{2} = 0.03M_{\odot}$,
$e_{2,i}=0.6$, $t_{\rm v,p} = 0.03$ yr. The values of $e_{1,i}$ are
chosen from 0.6 to 0.96 with a constant interval of 0.02, and
$\epsilon_i$, from 0.005 to 0.15 with a constant interval of 0.001.  The
final states are indicated by green crosses for Disrupted planets (TD),
black open squares for Non-migrating planets (NM), red filled circles
for Prograde Hot Jupiters (PHJ), and blue filled circles for Retrograde
Hot Jupiters(RHJ), respectively.  \label{fig:a2500}}
\end{center}
\end{figure}

\begin{figure}[h]
\begin{center}
\includegraphics[width=15cm]{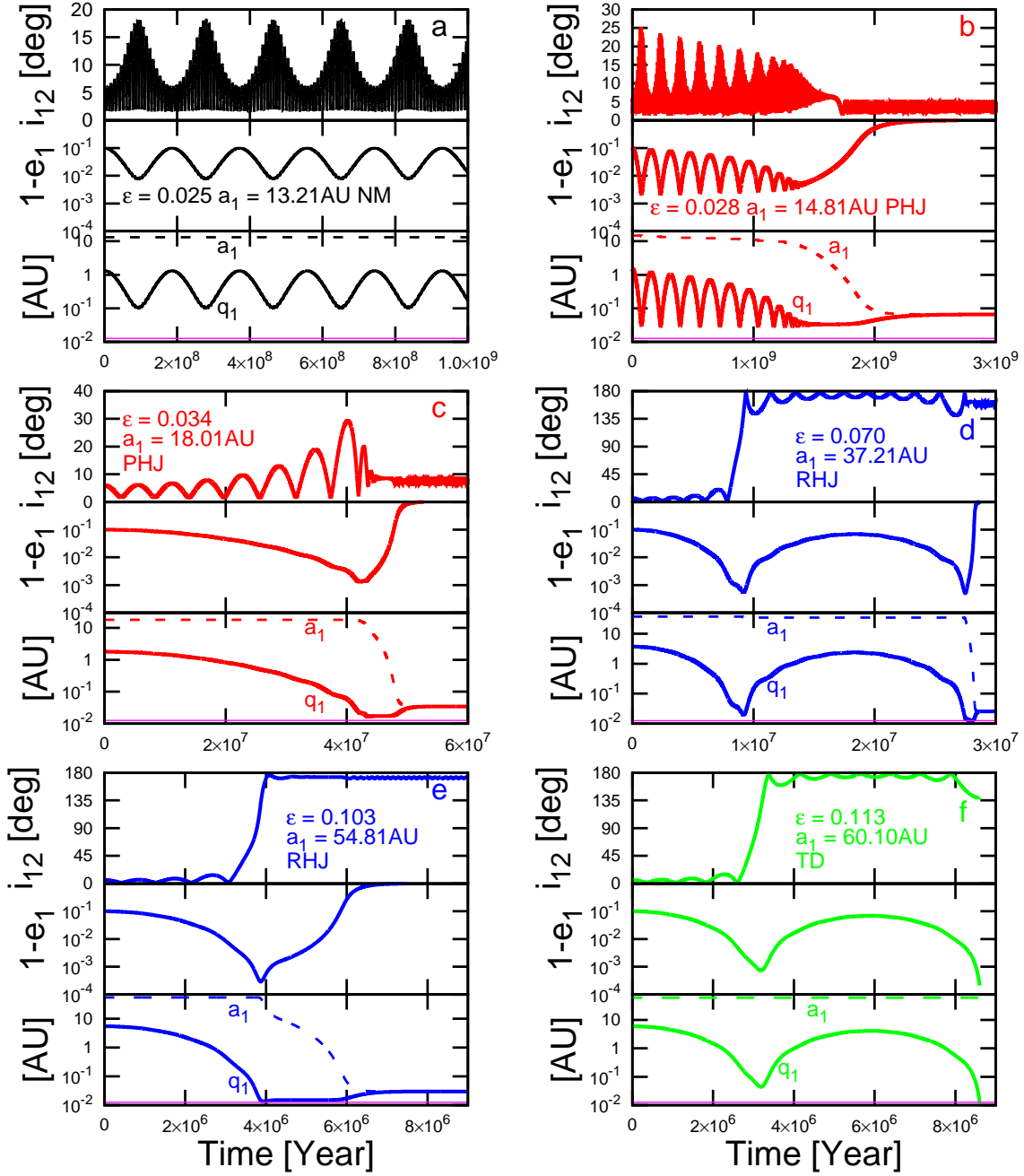} 
\caption{Evolution of our fiducial model with $e_{1,i} = 0.9$ for
different initial semi-major axis $a_{1,i}$.  The final outcomes,
Disrupted(TD), Non-migrating(NM), Prograde hot Jupiter(PHJ), and
Retrograde hot Jupiter(RHJ) are shown in green, black, red, and blue
line, respectively. For each time evolution, the evolution of $i_{12}$,
$e_{1}$, and $a_{1}$, $q_{1}$ are shown in the top, middle and bottom
panel, while $a_{1}$ is shown in dashed line, $i_{12}$, $e_{1}$, and
$q_{1}$ are shown in solid line, and Roche limit is shown in the bottom
panel with pink solid line, respectively. } \label{fig:samplee90}
\end{center}
\end{figure}

Figure \ref{fig:a2500} plots the final states of the inner planet in our
fiducial model for coplanar hierarchical triple systems.  In
this particular example, we first explore the range of
$0.005<\epsilon_i<0.15$ so as to make sure of the validity of the
analytical flip conditions, equation \ref{eq:flipcondition} by
\citet{Li2014} and equation \ref{eq:epsilon-oct} by
\citet{Petrovich2015b}.  Figure \ref{fig:a2500} clearly shows that the
region below those flip conditions agree with that of non-migrating
planets in our runs. So their conditions are accurate in distinguishing
the significant migration and non-migration boundary, even if they do
not necessarily lead to RHJs; see discussion below. Its most important
conclusion is that retrograde HJs are very difficult to form, despite
the fact that the plotted region of $e_{1,i}-\epsilon_{i}$ is chosen so
as to satisfy the flip condition (\ref{eq:flipcondition}) in the
massless limit; $\sim 90$\% of the systems are tidally disrupted, and
$\sim 10$\% survive as prograde HJs. The fraction of retrograde HJs
turn out to be less than 1\%.  Since this may be a rather unexpected
result, we plot the dynamical evolution of typical systems for
$e_{1,i}=0.9$ in six panels of Figure \ref{fig:samplee90}. While
we adopt 10 days as the spin rotation period of the inner planet
throughout the current analysis, it may be more relevant to use 10 hours
as in the case of Jupiter. In reality, however, the result turns out to
be fairly insensitive to the value as shown in \ref{sec:w10h} below.

The lower-limit of the analytical flip condition
(\ref{eq:flipcondition}) by \citet{Li2014}, $\epsilon_{\rm crit,i}$ is
a very good approximation for the {\it necessary condition}, but
obviously not a sufficient condition because it is
derived on the basis of orbital dynamics without short-range forces
effects. Our simulation shows that $\epsilon_{\rm crit,i}$ becomes
slightly larger, especially for large $e_{1,i}$(small $a_{1,i}$). The
detail of short-range forces effects is described in section
\ref{subsec:srfeffect}. One example of the behavior in the region
between the $\epsilon_{\rm crit,i}$ we adopted and the real flip
boundary including the short-range forces effects for $\epsilon_i=0.025
(a_{1,i}=13.21 {\rm AU})$ is illustrated in Figure
\ref{fig:samplee90}a. The system exhibits an oscillation both in $1-e_1$
and $i_{12}$, but the resulting pericenter distance $q_1$ is not small
enough for the tidal effect to operate. Thus the semi-major axis $a_{1}$
stays constant, and no significant migration occurs for
$10^{10}$yrs. All the other systems with $\epsilon_{i}<\epsilon_{\rm
crit, i}$, therefore not simulated in the present paper, show the same
behavior.

If $\epsilon_{i}$ is slightly larger than $\epsilon_{L}$, the amplitude of
oscillation in $1-e_1$ becomes larger as plotted in Figure
\ref{fig:samplee90}b for $\epsilon_i=0.028 (a_{1,i}=14.81 {\rm AU})$.
In this case, the maximum eccentricity reaches 0.998, rather than 0.990
in the non-migrating example. This large-amplitude oscillations allow
the inner planet reaching a minimum pericenter distance of
$a_{1}(1-e_{1}) \sim 0.03$ AU where tidal dissipation efficiently
extracts orbital energy, which results in gradual damping of $a_1$ at
each maximum eccentricity (minimum pericenter distance).  Thus PHJ
systems form via multiple close approaches within a typical timescale of
several $10^{9}$ yr. The example of Figure \ref{fig:samplee90}b results
in a HJ at $a \sim 0.065$ AU with $i_{12} \sim 4^{\circ}$. Indeed this
slow coplanar migration is systematically studied by
\citet{Petrovich2015b}, who proposes this as a potential path to PHJs,
and our results are in agreement with his proposal.

As $\epsilon_{i}$ increases further, the octupole potential starts to
dominate and drives $e_{1}$ very close to unity. At the same time, the
orbit flip happens if the dissipative tide is neglected. Along the line
of $e_{1} = 0.9$, we observe two continuous regions where PHJs form
($9.0\%$) via the coplanar-flip mechanism. One example in this region is
shown in Figure \ref{fig:samplee90}c for $\epsilon_i=0.034
(a_{1,i}=18.01 {\rm AU})$.  Its suggests that this path to PHJ happens
over a much smaller time-scale than that of Figure \ref{fig:samplee90}b;
note the different scales of time in each panel.  In this case, $1-e_1$
monotonically decreases and becomes close to $\sim 10^{-3}$ where the
tidal effect becomes important. Therefore in the middle of increasing
$i_{12}$, the system starts to be circularized and becomes PHJ with
$a_{1.f} \sim 0.035$ AU within $<10^7$ yrs. The mutual orbital
inclination oscillates with gradually increasingly the amplitude in the
range of $i_{12}\sim 0^{\circ}-30^{\circ}$, and then damps from $\sim
22^{\circ}$ to $\sim 9^{\circ}$ during the circularization stage. Since
the eccentricity increases until the end of orbit flip if no short-range
force effects are taken into account, such HJs have relatively low
$i_{12}$.  In total, the resulting PHJs (PHJ 9.0\%) are preferentially
located in the low $\epsilon$ region. Most of them are formed through
the coplanar-flip mechanism within a very short timescale ($\sim 10^{7}$
yr), while a few result from secular tidal damping via
eccentricity-inclination oscillation.

Beyond that value of $\epsilon_{i}$, the orbit of the inner planet is
indeed flipped, but the fate changes very sensitively due to the subtle
competition between the flipped condition and the tidal disruption as
illustrated in Figure \ref{fig:samplee90}d to f. As a result, the system
behavior looks chaotic, and there seems no systematic parameter region
for the formation of RHJ (see Figure \ref{fig:a2500}).

Evolution for the formation of RHJ similar to Figure
\ref{fig:samplee90}d and e happens only in a very narrow parameter
range; $\epsilon_i=0.070 (a_{1,i}=37.21 {\rm AU})$ and $\epsilon_i=0.103
(a_{1,i}=54.81 {\rm AU})$, respectively. The former is circularized at
the second closest point of $1-e_1$. The orbit suffers from tidal
circularization during an orbit flip process within a timescale of a few
$10^{7}$ yr. Since the eccentricity of the inner orbit increases in the
orbit flip stage, the system suffers from tidal circularization in the
beginning of the orbit flip stage in order not to be tidally disrupted.
Thus, this system ends with $i_{12,f}=162^{\circ}$, only slightly
smaller than the highest $i_{12}$ ever reached, $177^{\circ}$. While the
latter is circularized at the first closest point due to the stronger
perturbation of the outer body. The tidal circularization starts when
the orbit flip process is completed. Since the tidal circularization
does not modify $i_{12}$ significantly, $i_{12,f}$ remains almost
unchanged in the counter-orbiting regime, $172^{\circ}$ with $\pm
1^{\circ}$ oscillation. Such a high value of $i_{12}$ suggests that the
counter orbiting HJ can be formed via the coplanar-flip
mechanism, which supports the conclusion of \citet{Li2014}.

Figure \ref{fig:samplee90}f presents an example of a tidally disrupted
inner planet for $\epsilon_i=0.113 (a_{1,i}=60.10 {\rm AU})$.  Its
pericenter falls into the Roche limit at the second extreme eccentricity
approach when $1-e_1$ reaches $\sim 2\times10^{-4}$. Such a state is
preferentially found in systems in which the inner planet has a
relatively large semi-major axis, since the gravitationally interaction
between two orbits are stronger when the inner orbit reaches the extreme
eccentricity. The comparison among the panels d, e and f as well as
Figure \ref{fig:a2500} strongly indicates that the fate of the systems
is very sensitive to the parameters. Nevertheless the conclusion that
most of the systems satisfying the flip condition
(\ref{eq:flipcondition}) are tidally disrupted, instead of forming
counter-orbiting HJs, is quite general.

\clearpage

Since most of systems become disrupted via the orbital flip in our
simulation, the condition of forming retrograde or counter-orbiting HJs
is fairly fine-tuned. Considering the two successful examples of RHJs as
shown above, a subtle change of initial condition may singnificantly
modify the evolution and tidally disrupt the system as shown in Figure
\ref{fig:a2500}. So we may need to fine-tune the parameter sets in order
to successfully make RHJs, which seems to be unlikely.  Based on the low
ratio (RHJ 0.4\%) and such an uncertainty, it is difficult to form
retrograde or counter-orbiting HJs via the coplanar-flip mechanism.

Before moving to the next subsection, we would like to note that
there is an interesting pattern in Figure \ref{fig:a2500}; there are a
few branching structures in prograde hot Jupiters. These are more
significant in Figures \ref{fig:tvdm03} and \ref{fig:dism03}
below. Although we are not yet successful in explaining the behavior, we
suspect that they are related to some timescales in orbital
evolution. We hope to come back to the issue in our next paper.

\subsection{Effect of short-range forces} \label{subsec:srfeffect}

\citet{Liu2015} showed that the pericenter precessions due to
short-range force effects suppress the growth of eccentricity of the
inner planet, and reduce the flip region of $i_{12}$ for systems under
the Lidov-Kozai oscillation. In this subsection, we show that
the similar suppression works also for the near-coplanar triple systems.

A small area around the bottom-right region of Figure
\ref{fig:a2500} corresponds to non-migrating planets despite the fact
that they satisfy the flip criterion equation (\ref{eq:flipcondition})
initially. Indeed this comes from the short-range force effects. In
order to see their effects separately, we consider the NM
(non-migrating) planet example of Figure \ref{fig:samplee90}a
($\epsilon_i=0.025 (a_{1,i}=13.21 {\rm AU}), e_{1,i}=0.9$).

\begin{figure}[h]
\begin{center}
\includegraphics[width=14cm]{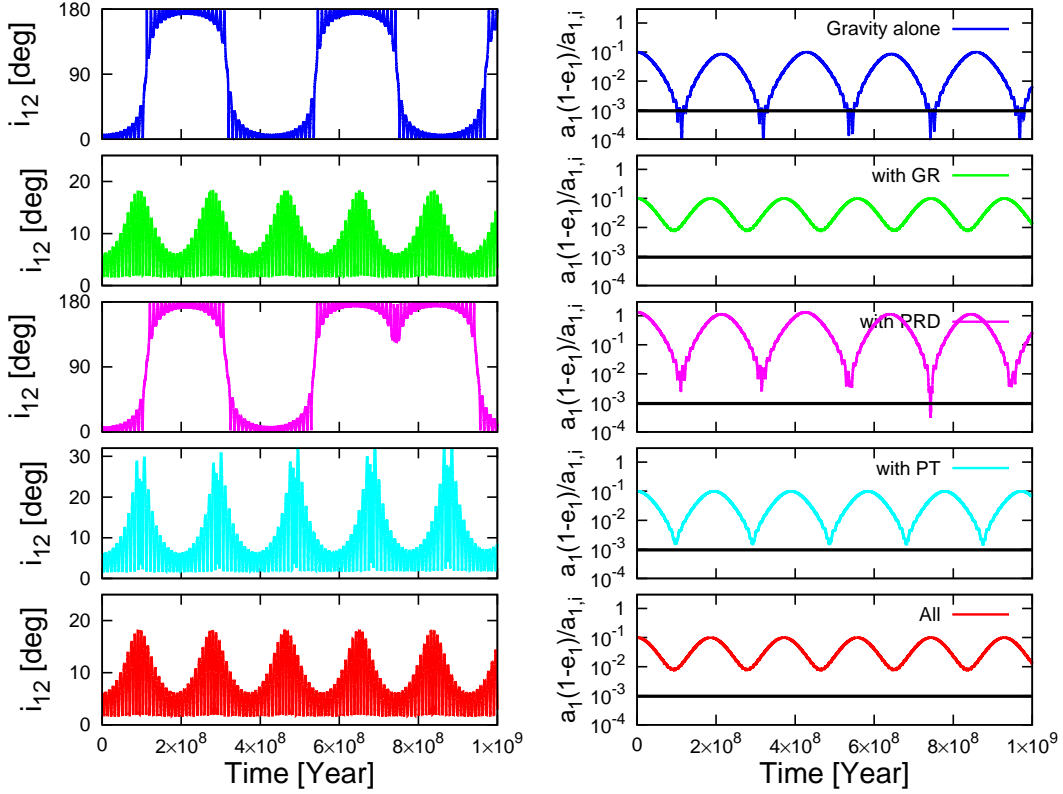} 
\caption{An illustrative example indicating the short-range force
effects. The initial condition of this example corresponds to that of
Figure \ref{fig:samplee90}a; $a_{1,i}=13.21$ AU ($\epsilon_{i}=0.025$),
and $e_{1,i}=0.9$.  Orbital evolution of $10^{9}$ yr with different
short-range force effects is plotted separately. From top to bottom, we
plot quadrupole and octupole gravitational force alone in blue, gravity
plus correction for general relativity (GR) in green, gravity plus
planetary rotational distortion (PRD) in magenta, gravity plus tides
(PT) in cyan, and finally gravity plus all the three short-range forces
(All) in red.  The black line corresponds to the Roche limit with
$f=2.7$.}  \label{fig:nmcheck}
\end{center}
\end{figure}

The left and right panels of Figure \ref{fig:nmcheck} plot the evolution
of the mutual orbital inclination, $i_{12}$, and the pericenter distance
of the inner planet in units of its initial semi-major axis. Since this
example corresponds to the NM case, the latter is almost equivalent to
$1-e_1$. We show the results for the secular orbital perturbation effect
alone, orbital and general relativistic (GR) correction, orbital and
planetary rotational distortion (PRD), orbital and planetary tide
(PT)\footnote{we include the central stellar tide and rotational
distortion as well in our simulation, but their effects are indeed
negligible.}, and orbital and all the short-range force effects, from
top to bottom.

As expected, the case without the short-ranges forces (top panels) flips
the orbital inclination each time $1-e_1$ becomes less than $\sim
10^{-3}$. The flip repeats periodically since no other dissipational
effects are included.  If only the PRD is included, the system still
shows the orbital flip, but the maximum value of $e_1$ is slightly
suppressed relative to the purely orbital case.

\clearpage

The precession due to PT could effectively limit the orbital flip with
maximum eccentricity less than $0.999$. On the other hand, the effect of
GR is very effective in suppressing the eccentricity; the maximum value
of $e_{1}$ under the GR correction barely reaches $\sim 0.99$. Thus the
system stays outside the tidal circularization region for $10^{10}$ yr,
and the PT never becomes important in reality, as shown in the bottom
panels of Figure \ref{fig:nmcheck}.

\begin{figure}[h]
\begin{center}
\includegraphics[width=8cm]{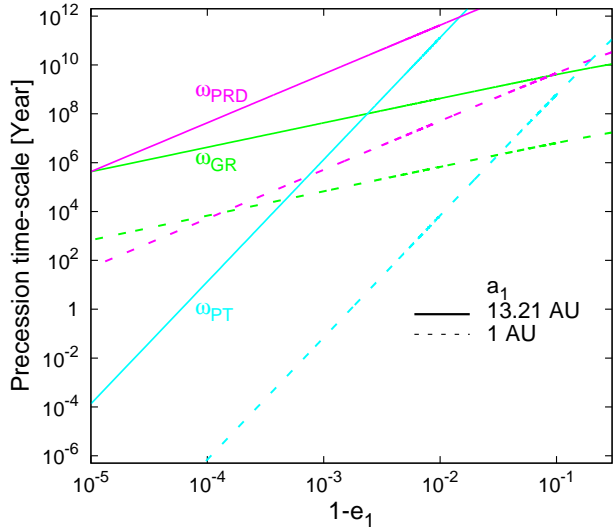} 
\caption{Analytical
precession time-scales for the three short-range forces on $\hat
{\mathbf e}_{1}$ as a function of $1- e_{1}$ (instead of
$1-e_{1}^{2}$). The solid and dashed lines correspond to $a_{1}=13.21
\rm AU$ (corresponding to Figure \ref{fig:nmcheck}) and $a_{1} = 1 \rm
AU$, respectively. The analytical expressions are explicitly given as
equations (\ref{eq:omegaGR}) $\sim$ (\ref{eq:omegaPRD}) in
\ref{sec:srf-eqs}.} \label{fig:srf-nmcheck}
\end{center}
\end{figure}

The above behavior can be understood by comparing the precession
time-scales of the Lenz vector $\hat{\mathbf e}_{1}$ for those
short-range forces, which we plot in Figure \ref{fig:srf-nmcheck} on the
basis of the expressions in \ref{sec:srf-eqs}.  Clearly the GR plays a
dominant role for $e_{1} < 0.995$, while PT becomes dominant for $e_{1}
> 0.995$; PRD is sub-dominant in either case. This is in good agreement
with our simulation result shown in Figure \ref{fig:nmcheck}, and
therefore the precession induced by the short-range forces, in
particular GR, prevents the orbital flip.  Since the short-range forces
become stronger for the smaller semi-major axis, the NM planets are
located around the high-$e_{1,i}$ and low-$\epsilon_{i}$ region.

\section{Dependence on the model parameters \label{sec:dependence}}

The previous section has presented the result for our fiducial model,
and discussed the dynamical behavior for several examples. Next we
consider the dependence of parameters employed in the fiducial model,
separately in each subsection below. The full list of different models is
summarized in Table \ref{tab:simparam}, and we plot the two models in
each subsection as examples. Since we already confirmed that planets with $\epsilon_i<\epsilon_{\rm crit}$ do not migrate in practice, we run the models for $\epsilon_{\rm crit}<\epsilon_i<0.15$ in what follows.

\subsection{Mass of the outer perturber \label{subsec:mdep}}

We adopt $m_2=0.03M_\odot$ as our fiducial value, but one might wonder
if the larger mass would be more relevant as (sub-)stellar perturbers.
While this sounds reasonable, the larger $m_2$ significantly increases
the tidal disruption ratio, and there is no chance to form  retrograde
planets in practice.

\begin{figure}[h]
\begin{center}
\includegraphics[width=15cm]{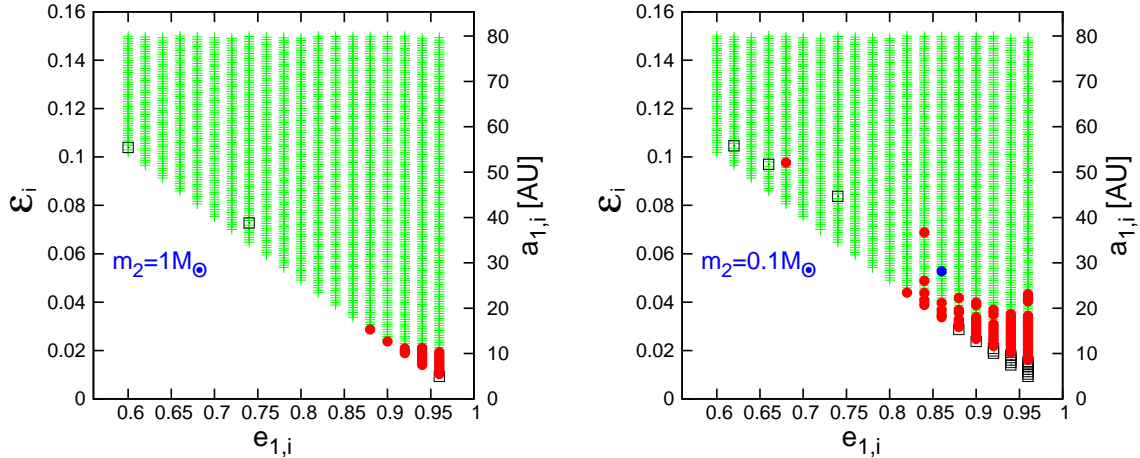} 
\caption{Final outcomes of m100 with  $m_{2}=1M_{\odot}$ (left) 
and m010  with $m_{2}=0.1M_{\odot}$ (right)
on $e_{1,i}-\epsilon_{i}$ plane.} \label{fig:m03mdep}
\end{center}
\end{figure}

This is clearly shown in the left and right panels of Figure
\ref{fig:m03mdep} for $m_2=1M_\odot$ and $0.1M_\odot$, respectively.
Since we focus on the parameter space satisfying the analytic flip
condition (\ref{eq:flipcondition}), the gravitational perturbation due
to the outer body is sufficiently strong to produce the orbital flip
potentially. Under such circumstances, the larger $m_2$ results in the
larger $e_1$ (extremely closer to unity) in which leads to the stronger
tidal effect. Therefore in order to survive the tidal disruption, the
inner planet should have the smaller $a_{1,i}$ for the larger
$m_2$. This is why the fractions of both PHJ and RHJ decreases as $m_2$
increases.

Thus it is very difficult to form RHJ via
the near-coplanar flip mechanism if the outer perturber has a stellar
mass $m_2>0.1M_{\odot}$. This is why we adopt $m_2=0.03M_\odot$ as our
fiducial value.

\subsection{Semi-major axis of the outer perturber \label{subsec:adep}}

\begin{figure}[h]
\begin{center}
\includegraphics[width=15cm]{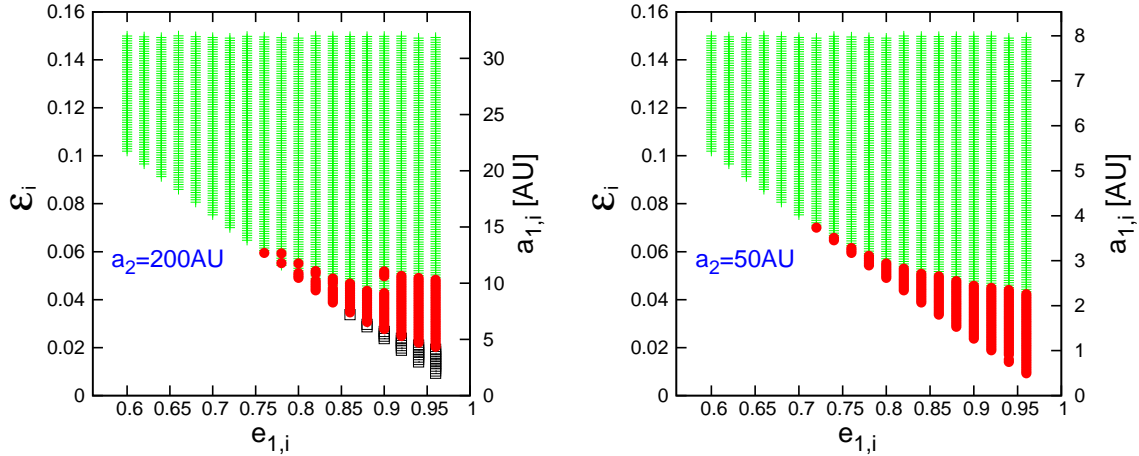} 
\caption{Final outcomes of a200 with  $a_{2}=200$AU (left) 
and a050 with $a_{2}=50$AU (right)
on $e_{1,i}-\epsilon_{i}$ plane.} \label{fig:m03ad}
\end{center}
\end{figure}

Consider next the dependence on $a_2$. Again the proper choice of this
parameter is not easy. If $a_2$ is larger, the near-coplanar
configuration is unlikely. On the other hand, the sub-stellar perturber
closer to the central star may be difficult to form either. As a
compromise, we select $a_{2}=500$, 200, 100 and $50$ AU in Table
\ref{tab:simparam} with $a_2=500$ AU being the fiducial value.
Figure \ref {fig:m03ad} presents the results for $a_2=200$AU and 50AU.
There are two important messages from Figure \ref {fig:m03ad}.  

First, RHJs do not form for $a_{2} \le 200$ AU.  In order to become a
RHJ, the inner planet needs to experience the orbital flip before the
tidal circularization. This prefers larger $a_{1,i}$ because the inner
planet suffers from less tidal dissipation before reaching the extreme
eccentricity for the orbital flip.  Even larger $a_{1,i}$, however,
results in stronger gravitational perturbation from the outer body, and
thus the inner planet is tidally disrupted. Due to that subtle
competition, RHJs in our fiducial model are confined in the narrow
region of $0.07<\epsilon_{i}<0.11$.  As $a_{2}$ decreases, the entire
system becomes more compact for the same value of $\epsilon_i$. Thus the
stronger gravitational perturbation of the outer body brings the inner
planet to the orbit within the Roche limit more easily because the
pericenter distance at the same maximum eccentricity is smaller.  This
is why RHJs disappear for the smaller $a_{2}$ models.  For the same
reason, PHJs are limited for the lower $\epsilon_{i}$ region.

Second, NM planet fraction drops as $a_{2}$ decreases; $1.8\%, 1.8\%,
1.1\%$, and $0.0\%$ for $a_{2} = 500, 200, 100$, and $50$ AU,
respectively.  In the fiducial model, short-range forces suppress the
the maximum eccentricity and the inner planet does not flip nor is
tidally circularized around the high-$e_{1,i}$ and very
low-$\epsilon_{i}$ region. The same value of $e_i$, however, corresponds
to the smaller $a_1$ for the smaller $a_{2}$ models. Thus the pericenter
distance for those systems becomes smaller, which enhances the tidal
dissipation and thus circularizes the orbit.  As a result, systems
gradually migrate and finally become PHJs via the secular
eccentricity-inclination oscillation as illustrated in Figure
\ref{fig:samplee90}b.
In any case, the formation of RHJs is more difficult for the smaller
$a_2$ than the fiducial model.

\subsection{Eccentricity of the outer perturber \label{subsec:edep}}

\begin{figure}[tbh]
\begin{center}
\includegraphics[width=15cm]{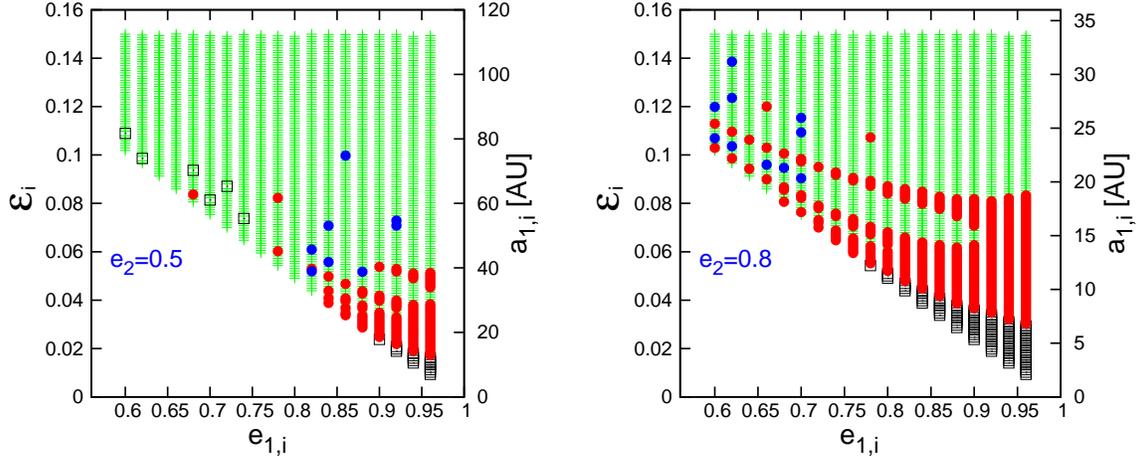} 
\caption{Final outcomes of e05 with  $e_{2,i}=0.5$ (left) 
and e08 with $e_{2,i}=0.8$ (right)
on $e_{1,i}-\epsilon_{i}$ plane.} \label{fig:m03edep}
\end{center}
\end{figure}

Sub-stellar perturbers may exhibit a broad range of eccentricity, and we
run six simulation sets with $e_{2,i}$ = 0.3, 0.4, 0.5, 0.6 (fiducial),
0.7, and 0.8, and two examples out of those models are plotted in Figure
\ref{fig:m03edep}.
We find that the fraction of PHJs monotonically increases for the more
eccentric outer perturber.  PHJs tend to form preferentially in low
$a_{1,i}$ where tidal dissipation becomes effective.  Since we consider
the same range of $\epsilon_i$ for all the models, the corresponding
value of $a_{1,i}$ for the same $\epsilon_i$ becomes smaller as $e_{2,i}$
increases. Thus the dependence of the fraction on $e_2$ is mainly due to
the scaling.
While RHJs are very rare, their fraction also increases slightly as
$e_2$, but it would be mainly due to the scaling of $a_{1,i}$ with
respect to $\epsilon_i$.

\subsection{Mutual orbital inclination of the inner and outer orbits 
\label{subsec:i12}}

The initial orbits of the inner and outer bodies are naturally expected
to be inclined to some extent. While our fiducial model adopts
$i_{12,i}=6^\circ$, we examine more inclined cases of
$i_{12,i}=15^\circ$ and $30^\circ$ as well as an idealized coplanar case
($i_{12,i}=0^\circ$). The Lidov-Kozai mechanism starts to work for more
inclined cases, and we do not consider here because the orbital flip
does not happen in those cases as mentioned in Introduction.

\bigskip

\begin{figure}[tbh]
\begin{center}
\includegraphics[width=15cm]{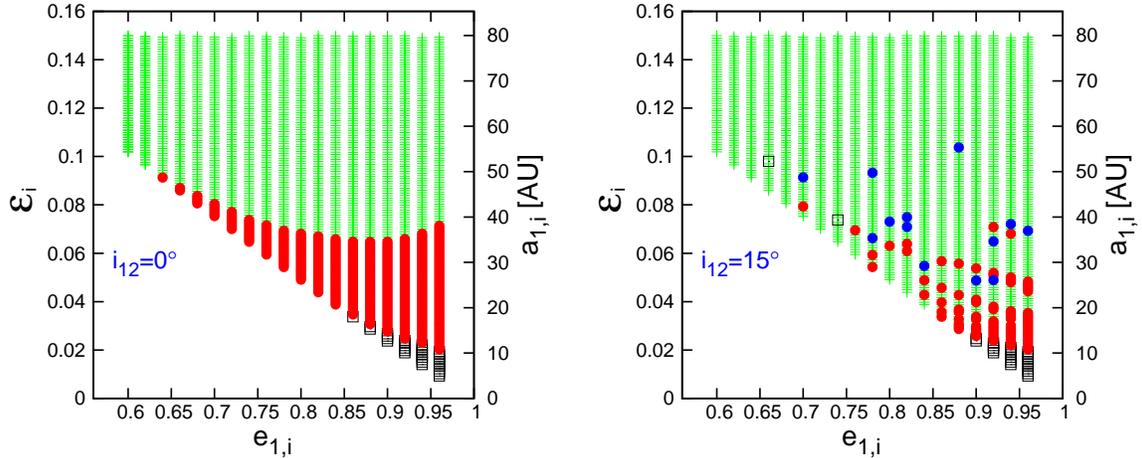} 
\caption{Final outcomes of i00 with  $i_{12,i}=0$ (left) 
and i15 with $i_{12,i}=15^\circ$ (right)
on $e_{1,i}-\epsilon_{i}$ plane.} 
\label{fig:idepm03}
\end{center}
\end{figure}

The left and right panels of Figure \ref{fig:idepm03} present the
results for $i_{12} = 0$ and $i_{12} = 15^{\circ}$.  In the exact
coplanar case, the net force normal to the orbital plane always
vanishes, and the orbits cannot flip. Thus RHJs cannot form, but PHJs
can.

As $i_{12,i}$ increases, the fraction of PHJs decreases monotonically,
and they are confined around the narrow region with high-$e_{1,i}$ and
low $\epsilon_{i}$. In rare cases, RHJs form in a scatter manner over
the on $e_{1,i}-\epsilon_{i}$ plane, probably due to the chaotic nature
of the system.

\subsection{Viscous time-scale of the inner planet \label{subsec:tdep}}

Unfortunately it is well known that the viscous time-scale of planets,
$t_{\rm v,p}$, (equivalently, the tidal delay time and tidal quality
factor) is the most uncertain parameter in the equilibrium tidal theory.
The observational data for the Jupiter -- Io system put an empirical
lower limit on that of Jupiter as $t_{\rm v,J}>15$ yr.  On the other
hand, \citet{Socrates2012} stated that $t_{\rm v,p} < 1.5$ yr for an
initially highly eccentric planetary orbit with semi-major axis of $\sim
5$ AU to be circularized into $<0.06$ AU within 10 Gyr.  They argued that
the discrepancy between their upper limit and the empirical lower limit
for Jupiter should not be taken seriously given various theoretical
uncertainties concerning the tidal dissipation model and diversities of
the physical properties of the exoplanets.

\bigskip

\begin{figure}[bh]
\begin{center}
\includegraphics[width=15cm]{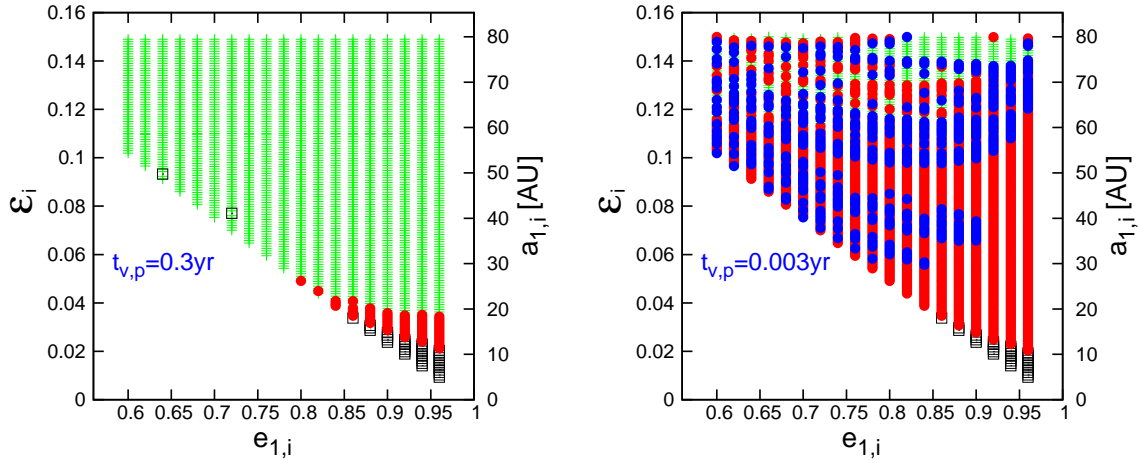} 
\caption{Final outcomes of t03000 with  $t_{\rm v,p}=0.3$yr (left) 
and t00030 with  $t_{\rm v,p}=0.003$yr (right) 
on $e_{1,i}-\epsilon_{i}$ plane.} \label{fig:tvdm03}
\end{center}
\end{figure}

For instance, more recent work by \citet{Storch2014} examined a
possibility of tidal dissipation in solid cores of giant planets, and
claimed that tidal dissipation in the core can reconcile the Jupiter-Io
tidal constraint and very efficient high-eccentricity migration
simultaneously.

Given a somewhat confusing situation, we decided to adopt $t_{\rm v,p} =
0.03$ as our fiducial value, simply following \citet{Li2014}.  Our
purpose of the present paper is not to find a suitable value for $t_{\rm
v,p}$ but to understand the role of $t_{\rm v,p}$ in the orbit flip of
near-coplanar triple systems.  Thus we examine the other three cases
with $t_{\rm v,p}$ = 0.3, 0.003, and 0.0003 yr as well.

The results are plotted in Figure \ref{fig:tvdm03}. As expected, the
fate of the inner planet is very sensitive to the very uncertain value
of $t_{\rm v,p}$.  When $t_{\rm v,p}$ is smaller, the tide on the planet
becomes stronger and the planet suffers from very efficient
circularization even at a larger pericenter distance. Thus the majority
of the tidally disrupted planets for $t_{\rm v,p}=0.3$ yr survive as
PHJs and RHJs for $t_{\rm v,p}=0.003$yr.¡¡The lower-right region of
Figure \ref{fig:tvdm03} corresponds to planets at a relatively larger
pericenter distance, and thus insensitive to the value of $t_{\rm v,p}$.

Of course, the value of $t_{\rm v,p}=0.003$yr is very extreme and
unrealistic; even the paucity of the observed RHJs is inconsistent with
the choice. Nevertheless Figure \ref{fig:tvdm03} clearly illustrates
that the uncertainty of the tidal dissipation model is the key to
understanding the formation and dynamical evolution of HJs in general.

\subsection{The proportional constant for the Roche limit 
\label{sec:rochedep}}

Finally we consider the criterion of the tidal disruption itself.  As
discussed in \S \ref{subsec:model}, the proportional factor $f$ of the
Roche limit in equation (\ref{eq:roche-limit}) is not precisely
determined.  While we adopt $f=2.7$ following \citet{Guillochon2011}
from hydrodynamical simulations, $f=2.16$ is reported by
\citet{Faber2005} and $f=1.66$ is adopted in simulations by
\citet{Naoz2012}.

\bigskip

\begin{figure}[h]
\begin{center}
\includegraphics[width=15cm]{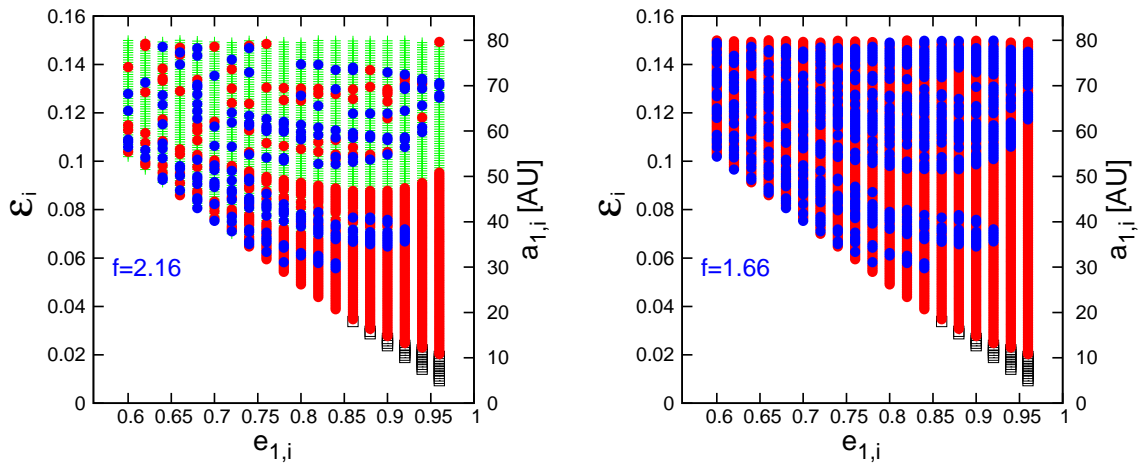} 
\caption{Final outcomes of f216 with  $f=2.16$ (left) 
and f166 with  $f=1.66$ (right) 
on $e_{1,i}-\epsilon_{i}$ plane.}\label{fig:dism03}
\end{center}
\end{figure}

As shown in the previous subsection, the efficiency of the tidal
disruption is the most important in determining the fate of the inner
planet.  Thus we plot the cases of $f=2.16$ and $f=1.66$ in the left and
right panels of Figure \ref{fig:dism03}, respectively.

Similarly to Figure \ref{fig:tvdm03}, short-range forces are effective
and suppress the growth of the eccentricity of the inner planet in the
lower-right region of Figure \ref{fig:dism03}.  Thus the pericenter
distance of the inner planets around the region is larger than $R_{\rm
roche}$ in any case, and the fudge factor $f$ hardly changes the
evolution of those planets.

On the other hand, tidally disrupted planets in our fiducial model are
sensitive to the value of $f$. As is clear from Figure \ref{fig:dism03},
those planets turn out to survive as PHJs and RHJs for the smaller value
of $f$, and there are no tidally disrupted planets for $f=1.66$. Indeed
the result with $f=1.66$ is already virtually indistinguishable with the
case where the tidal disruption happens only when the inner planet falls
into the central star.

\section{Spin-orbit angle distribution \label{sec:spin-orbit}}

\begin{figure}[tbh]
\begin{center}
\includegraphics[width=14cm]{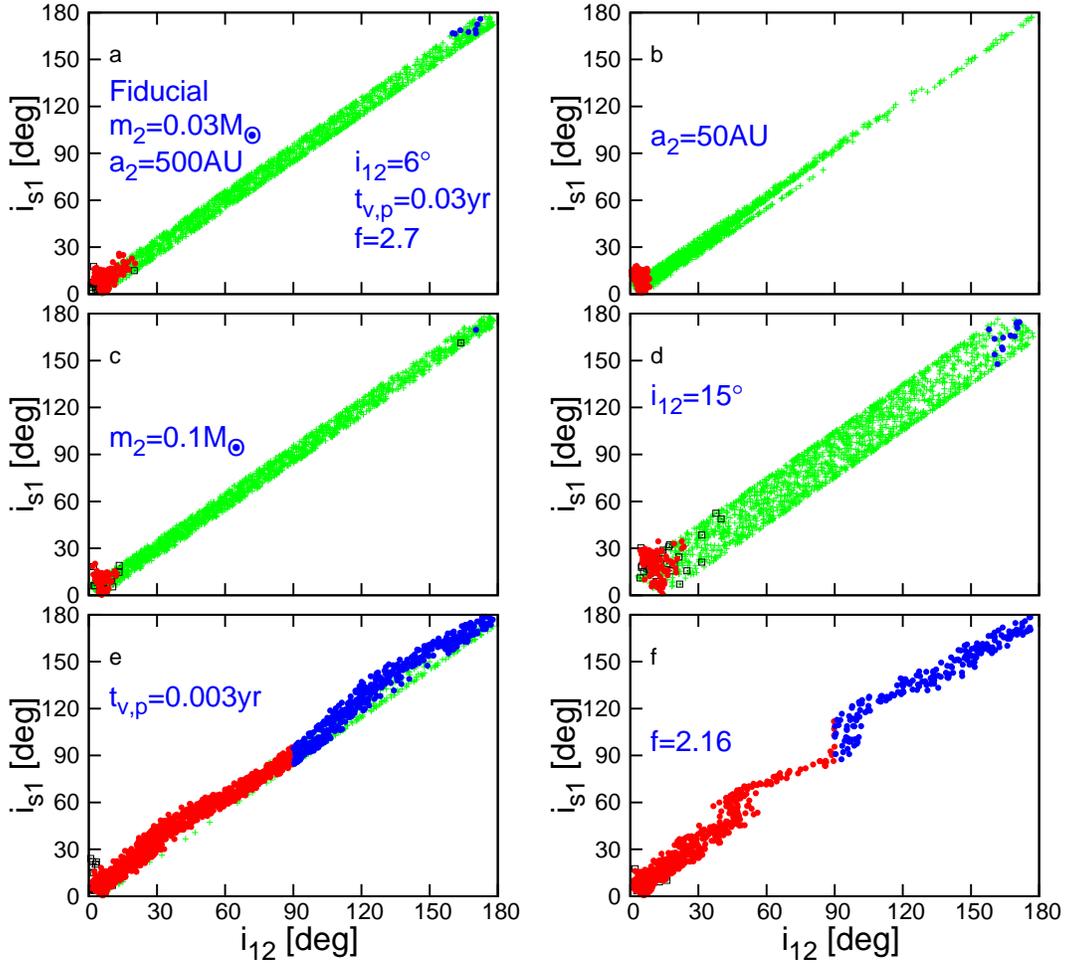} \caption{Orbital mutual
orbital inclination against the spin-orbit angle between the central
star and the inner planet.  The different colors indicate the different
final outcomes of the inner planet; NM (black), PHJ (red), RHJ (blue),
and TD (green).}  \label{fig:i12is1}
\end{center}
\end{figure}

So far we have classified the survived HJs into prograde or retrograde
according to the mutual orbital inclination angle $i_{12}$ of the inner
and outer orbits, i.e., $i_{12}<90^\circ$ or $>90^\circ$, respectively.
In reality, however, $i_{12}$ cannot be measured directly since
the possible outer perturbers of the observed HJs are hardly
identified. Thus observationally the distinction between prograde and
retrograde HJs is made from the the value of $\lambda$, the
sky-projected angle of $i_{s1}$, obtained from the
Rossiter-McLaughlin effect. Since our current simulation runs solve the
evolution of the stellar spin axis as well, we can address the validity
of a somewhat conventional assumption of $i_{12} = i_{s1}$.  The result
is plotted in Figure \ref{fig:i12is1}, which basically confirms that
$i_{12}$ can be used as a proxy for $i_{s1}$ as long as the stellar
spin vector is completely aligned with the orbital angular momentum
vector of the inner planet initially ($i_{s1,i}=0$) as we adopted in the
present runs.

\begin{figure}[tbh]
\begin{center}
\includegraphics[width=14cm]{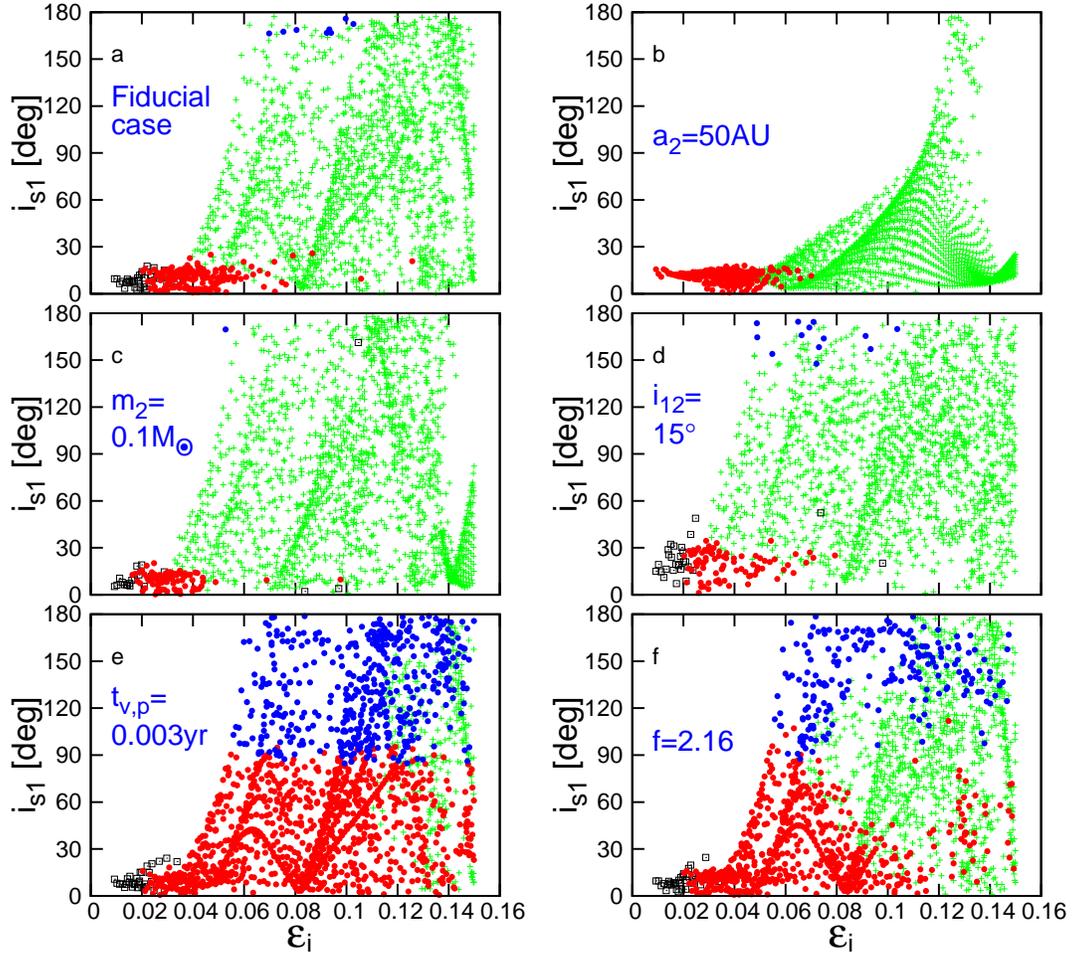}
\caption{Spin-orbit angles $i_{s1}$ for our models; a: fiducial, b: a50,
c: m01, d: i15, e: t00030, and f: f216.  The different colors indicate
the different final outcomes of the inner planet; NM (black), PHJ (red),
RHJ (blue), and TD (green).}  \label{fig:spinorbitangledis}
\end{center}
\end{figure}

Now we show the distribution of $i_{s1}$ in Figure
  \ref{fig:spinorbitangledis}. These plots indicate that PHJs and RHJs
  in our simulations correspond almost exclusively to well-aligned
  ($i_{12} < 20^\circ$) and counter-orbiting ($180^\circ- i_{12} <
  20^\circ$) planets. This is not the case, however, for models with
  very strong tidal interaction (t00030 and f216), which exhibit a very
  broad distribution of $i_{12}$ and thus of $i_{s1}$.

 In the coplanar-flip mechanism, the planetary orbit suffers
from tidal circularization after the orbit flip. Thus the system ends up
with PHJs if the tidal circularization happens before the orbit flip,
and RHJs if the orbit flip occurs before the circularization.
 On the other hand, the tidally disrupted planets have a
very broad distribution of $i_{s1}$ that we define at the epoch when the
pericenter distance of the inner planet reaches the Roche limit. The
result implies that those planets fall into the Roche limit in a very
short time-scale less than that of the orbit flip.

\clearpage

\clearpage

\section{Summary and discussion \label{sec:summary}}

 The observation of the Rossiter-McLaughlin effect has revealed a dozen
of possible retrograde planets, which already has challenged the
conventional theory of planet formation.  Although there exists no
reliable candidate (yet), the presence of counter-orbiting planets would
have an even stronger impact on the formation theory; somewhat
conventional planetary migration scenarios including disk-planet
interaction, planet-planet scattering, and the Lidov-Kozai migration are
successful in producing retrograde planets, but fail to explain the
counter-orbiting planets in general.

An interesting and attractive possibility is based on the extreme
eccentricity evolution expected for the near-coplanar hierarchical
triple system. Indeed \citet{Li2014} and \citet{Petrovich2015b} derived
an analytical condition for the orbital flip of the inner planet, which
holds for the massless limit of the inner planet under the quadrupole
and octupole gravitational potentials of the outer perturber but
neglecting the short-range forces (GR, star and inner planetary tide, and
rotational distortion) .

 In the present paper, we have performed a series of systematic
simulations for the sub-stellar outer perturber case, including the
short-range forces and examined in detail the condition for the orbital
flip in a more realistic situation.

Our main findings are summarized as follows;

1) Most of the near-coplanar hierarchical triple systems that satisfy
the analytical flip condition do not produce counter-orbiting
planets. Instead, the inner planets in those systems are tidally
disrupted.  A small fraction of the systems end up with the prograde Hot
Jupiters, and very few retrograde Hot Jupiters are produced.  Systems
that do not satisfy the analytical flip condition do not exhibit any
significant migration of the inner planet.

2) The break-down of the the analytical flip condition is due to the
short-range forces, which suppresses the extreme eccentricity evolution
of the inner planet that is required for the orbital flip.

3) The results are almost independent of the model parameters, and thus
fairly generic unless unrealistically strong tidal effect is assumed.

4) The mutual orbital inclination angle between the inner planet and
outer perturber, and the spin-orbit angle between the central star and
the inner planet are almost the same.  Their distribution for the
survived Hot Jupiters is bimodal; $\sim0^{\circ}-20^{\circ}$ for prograde,
and $\sim160^{\circ}-180^{\circ}$ for retrograde planets, and virtually
nothing in-between.

Our simulation runs span the parameter space that satisfy the analytical
flip condition, and more importantly uniformly sample the $\epsilon_{1,i}$
-- $e_{1,i}$ plane without assuming any prior distribution for their
realistic values. Therefore predicted statistics for the fate of the
inner planet under such configurations are significantly biased.  Having
emphasized such warnings, however, it might be instructive to present
some statistics simply illustrating the difficulty of forming the
counter-orbiting planets in the near-coplanar hierarchical triple
system.

\begin{figure}[h]
\begin{center}
\includegraphics[width=15cm]{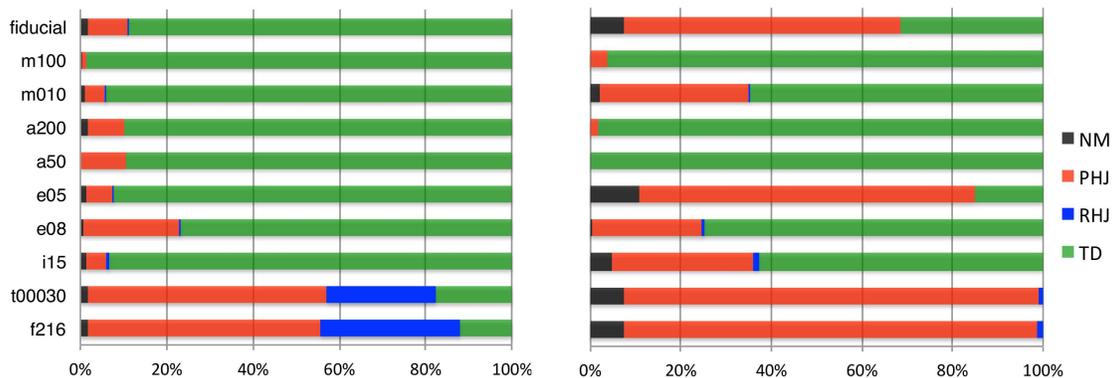} 
\caption{The fraction
of the final outcome of the inner planets.  Left panel: All simulation
runs. Right panel: $10 \rm AU < a_{1,i} < 30 \rm AU$.}
\label{fig:histo}
\end{center}
\end{figure}

Figure \ref{fig:histo} plots the fraction of four different final
outcomes of the inner planet; NM (non-migrating planet) in black, PHJ
(prograde Hot Jupiter)) in red, RHJ (retrograde Hot Jupiter) in blue,
and TD (tidally disrupted planet) in green.  The left panel corresponds
to a number fraction of each fate simply from the numbers out of 1800
runs for each model summarized in Table \ref{tab:simparam}. The right
panel is computed from their sub-sample with $10 {\rm AU} <a_{1,i}<30
{\rm AU}$ so as to sample the $a_{1,i}$ -- $e_{1,i}$ plane assuming the
eccentric inner gas giant planets orbiting at reasonable distances from
the central star just for the comparison purpose. 

In any case, our basic conclusion remains the same even if the
statistics shown here just for example may be highly biased; it is very
difficult to produce the retrograde planet in the present scenario,
while some fraction of prograde Hot Jupiter might have formed through
this channel.  This implies that the formation of counter-orbiting
planets imposes an even more serious challenge for the theory. Instead,
it could be simply the case that counter-orbiting planet candidates with
the projected spin-orbit angle $\lambda \approx 180^\circ$ are mildly
misaligned with their true spin-orbit angles $\psi$ being much less than
$180^\circ$ as suggested for HAT-P-7b \citep{Benomar2014}. In this
respect, future observational search for the counter-orbiting planets
combined with the Rossiter-McLaughlin effect and asteroseismology
continues to be important, and hopefully will bring an exciting puzzle
for planet formation.

Finally we note that the presence of numerous tidally disrupted planets
is not specific to the near-coplanar hierarchical triple systems, but a
fairly generic outcome in planetary migration models and in spin-orbit
realignment models \citep{Lai2012,Rogers2013,Xue2014,LiWinn2015}. Thus it
is of vital importance to look for possible signatures of such tidal
disruption events observationally. Indeed recent studies for the
determination of the orbital decay rate \citep{Jiang2015} and for the
unsual photometric signals in KIC 8462852
\citep{Boyajian2015,Bodman2015}, for instance, are closely related to
such an important direction.

\acknowledgments

We thank Shoya Kamiaka and Kento Masuda for useful discussions.
We are also grateful to an anonymous referee for several important
  suggestions that improved our earlier manuscript. This research is
supported by the Grant-in Aid for Scientific Research by Japan Society
of Promotion of Science No. 24340035.

\appendix
\renewcommand\thesection{\appendixname~\Alph{section}}
\renewcommand\theequation{\Alph{section}.\arabic{equation}}

\section{Basic equations for secular evolution
\label{sec:correia-eqs}}

Just for self-containedness, we write the secular equations of motion
used in the present paper for a hierarchical triple system.  We consider
gravitational interaction up to the octupole expansion of the outer body
as described by \citet{Liu2015}. In addition, we include the general
relativistic correction, the spin effect of the central star and the
inner planet, and tidal effect following \citet{Correia2011}. In
addition, we incorporate the damping of the stellar spin due to magnetic
braking following \citet{Barker2009}.

The subscripts $0$, $1$, and $2$ distinguish the quantities for the
central star, the inner planet and the outer perturber, respectively.
The mass and radius of those objects are denoted by $m$ and $R$.
The spin rate $\omega_{i}$
and gravity coefficients $J_{2_{i}}$ for the star ($i=0$) and inner
planet ($i=1$) are written as
\begin{equation} 
 J_{2_{i}} =k_{2_{i}}\frac{\omega_{i}^{2}R_{i}^{3}}{3Gm_{i}},
\end{equation}
where $k_{2_{i}}$ is the second Love number that characterizes the
deformation property of each body.

All the equations are written in Jacobi coordinates with $\mathbf r_{1}$
being the relative position from $m_{0}$ to $m_{1}$, and $a$ and $e$ are
semi-major axis and eccentricity, respectively. Then, the evolution of
spin and orbit can be tracked in the octupole approximation by three
parameters; spin angular momentum:
\begin{equation} 
 \mathbf L_{i}=C_{i}\omega_{i}\hat{\mathbf s}_{i},
\end{equation}
where $\hat{\mathbf s}_{i}$ is the unit vector of $\hat{\mathbf L}_{i}$ and
$C_{i}$ is the principal moment of inertia, the orbital angular momentum:
\begin{equation} 
\mathbf G_{i}=\beta_{i}
\sqrt{\mu_{i}a_{i}(1-e_{i}^{2})}\hat{\mathbf k}_{i},
\end{equation}
where $\hat{\mathbf k}_{i}$ is the unit vector of $\hat{\mathbf G}_{i}$
with $\beta_{1}=m_{0}m_{1}/(m_{0}+m_{1})$,
$\beta_{2}=(m_{0}+m_{1})m_{2}/(m_{0}+m_{1}+m_{2})$,
$\mu_{1}=G(m_{0}+m_{1})$ and $\mu_{2}=G(m_{0}+m_{1}+m_{2})$, and
finally the Lenz vector:
\begin{equation} 
\mathbf e_{1}=\frac{(\dot{\mathbf r}_{1}\times \mathbf
 G_{1})}{\beta_{1}\mu_{1}}-\frac{\mathbf r_{1}}{r_{1}}.
\end{equation}

We define direction angles as
\begin{equation} 
\cos\theta_{i}=\hat{\mathbf s}_{i}\cdot \hat{\mathbf k}_{1}, \qquad
\cos\epsilon_{i}=\hat{\mathbf s}_{i}\cdot \hat{\mathbf k}_{2}, \qquad
\cos i_{12} =\hat{\mathbf k}_{1}\cdot \hat{\mathbf k}_{2}, 
\end{equation}
where $\theta_{i}$ is the angle between the spin of the $i-$th body (in
the main text we use $i_{s1}$ to denote $\theta_0$),
$\hat{\mathbf s}_{i}$ and inner orbit, $\hat{\mathbf k}_{1}$,
$\epsilon_{i}$ is the angle between the spin of the $i-$th body,
$\hat{\mathbf s}_{i}$, and outer orbit $\hat{\mathbf k}_{2}$, and $i_{12}$ is
inclination between two orbits. 

Averaging the equations of motion over the mean anomalies of the inner
and outer bodies, we obtain the following equations for the conservative
motion:
\begin{eqnarray} 
\dot{\mathbf G}_{1}
&=&-\gamma(1-e_{1}^{2})\cos i_{12} \hat{\mathbf k}_{2}\times \hat{\mathbf
k}_{1} + 5\gamma(\mathbf e_{1}\cdot 
\hat{\mathbf k}_{2})\hat{\mathbf k}_{2}\times \mathbf e_{1} \nonumber\\
&&- \frac{25}{16}\epsilon_{\rm oct}\gamma
\Bigg\{\left[2(1-e_{1}^{2})[(\mathbf e_{1}\cdot \hat{\mathbf e}_{2})\cos i_{12}
+(\mathbf e_{1}\cdot \hat{\mathbf k}_{2})(\hat{\mathbf k_{1}}\cdot 
\hat{\mathbf e}_{2})]\hat{\mathbf k}_{1} \right. \nonumber\\
&& \left. +2[(1-e_{1}^{2})(\hat{\mathbf k_{1}}\cdot \hat{\mathbf e}_{2})\cos i_{12}-7(\mathbf e_{1}\cdot \hat{\mathbf k}_{2})(\mathbf e_{1}\cdot \hat{\mathbf k}_{2})]\mathbf e_{1}\right]\times \ \hat{\mathbf k}_{2}\nonumber\\
&& +\left[2(1-e_{1}^{2})(\mathbf e_{1}\cdot \hat{\mathbf k}_{2})\cos i_{12} \hat{\mathbf k}_{1}+[\frac{8}{5}\mathbf e_{1}^{2}-\frac{1}{5}\right. \nonumber\\
&&\left. -7(\mathbf e_{1}\cdot \hat{\mathbf k}_{2})^{2} + (1-e_{1}^{2})\cos^{2}i_{12}]\mathbf e_{1}\right]\times \hat{\mathbf e_{2}}\Bigg\} \nonumber\\
&& -\sum_{i}\alpha_{1i}\cos\theta_{i} \hat{\mathbf
 s}_{i}\times\hat{\mathbf k}_{1},
\end{eqnarray}
\begin{eqnarray} 
\dot{\mathbf G}_{2}&=&-\gamma(1-e_{1}^{2})\cos i_{12} \hat{\mathbf k}_{1}\times \hat{\mathbf k}_{2} + 5\gamma(\mathbf e_{1}\cdot \hat{\mathbf k}_{2})\mathbf e_{1}\times \hat{\mathbf k}_{2} \nonumber\\
&&+ \frac{25}{16}\epsilon_{\rm oct}\gamma
\Bigg\{\left[2(1-e_{1}^{2})[(\mathbf e_{1}\cdot \hat{\mathbf e}_{2})\cos i_{12}
+(\mathbf e_{1}\cdot \hat{\mathbf k}_{2})(\hat{\mathbf k_{1}}\cdot 
\hat{\mathbf e}_{2})]\hat{\mathbf k}_{1} \right. \nonumber\\
&& \left. +2[(1-e_{1}^{2})(\hat{\mathbf k_{1}}\cdot \hat{\mathbf e}_{2})\cos i_{12}-7(\mathbf e_{1}\cdot \hat{\mathbf k}_{2})(\mathbf e_{1}\cdot \hat{\mathbf k}_{2})]\mathbf e_{1}\right]\times \ \hat{\mathbf k}_{2}\nonumber\\
&& +\left[2(1-e_{1}^{2})(\mathbf e_{1}\cdot \hat{\mathbf k}_{2})\cos i_{12} \hat{\mathbf k}_{1}+[\frac{8}{5}\mathbf e_{1}^{2}-\frac{1}{5}\right. \nonumber\\
&&\left. -7(\mathbf e_{1}\cdot \hat{\mathbf k}_{2})^{2} + (1-e_{1}^{2})\cos^{2}i_{12}]\mathbf e_{1}\right]\times \hat{\mathbf e_{2}}\Bigg\} \nonumber\\
&& - \sum_{i}\alpha_{2i}\cos\epsilon_{i} \hat{\mathbf
 s}_{i}\times\hat{\mathbf k}_{2},
\end{eqnarray}
\begin{eqnarray} 
\dot{\mathbf e}_{1}&=&-\frac{\gamma(1-e_{1}^{2})}{||\mathbf G_{1}||}\left[\cos i_{12} \hat{\mathbf k}_{2}\times \mathbf e_{1} -2\hat{k}_{1}\times \mathbf e_{1} - 5( \mathbf e_{1}\cdot \hat{\mathbf k}_{2})\hat{\mathbf k}_{2}\times \hat{\mathbf k}_{1}\right]\nonumber\\
&& - \frac{25}{16}\epsilon_{\rm oct}\gamma\Bigg\{2\sqrt{(1-e_{1}^{2})}\left[(\mathbf e_{1}\cdot \hat{\mathbf k}_{2})\cos i_{12}\mathbf e_{1} \right. \nonumber\\
&& \left. + [\frac{8}{5}\mathbf e_{1}^{2}-\frac{1}{5}  -7(\mathbf e_{1}\cdot \hat{\mathbf k}_{2})^{2} + (1-e_{1}^{2})\cos^{2} i_{12}]\hat{\mathbf k}_{1}\right]\times \hat{\mathbf e_{2}}\nonumber\\
&& + 2\sqrt{(1-e_{1}^{2})}\left[[(\mathbf e_{1}\cdot \hat{\mathbf e}_{2})\cos i_{12}+(\mathbf e_{1}\cdot \hat{\mathbf k}_{2})(\hat{\mathbf k_{1}}\cdot \hat{\mathbf e}_{2})]\mathbf e_{1} \right. \nonumber\\
&& \left. +[(\hat{\mathbf k_{1}}\cdot \hat{\mathbf e}_{2})\cos i_{12}-7(\mathbf e_{1}\cdot \hat{\mathbf k}_{2})(\mathbf e_{1}\cdot \hat{\mathbf e}_{2})] \hat{\mathbf k}_{1}\right]\times \hat{\mathbf k}_{2}+\frac{16}{5}(\mathbf e_{1}\cdot \hat{\mathbf e}_{2})\sqrt{(1-e_{1}^{2})}\mathbf k_{1}\times \mathbf e_{1}\Bigg\}\nonumber\\
&& - \sum_{i}\frac{\alpha_{1i}}{||\mathbf G_{1}||}\left[\cos\theta_{i}
 \hat{\mathbf s}_{i}\times \mathbf
 e_{1}+\frac{1}{2}(1-5\cos^{2}\theta_{i})\hat{\mathbf k}_{1}\times
 \mathbf e_{1}\right] ,
\end{eqnarray}
\begin{eqnarray} 
\dot{\mathbf e}_{2}&=&\frac{\gamma}{\sqrt{1-e_{2}^{2}}} \frac{\beta_{1}\sqrt{\mu_{1}a_{1}}}{\beta_{2}\sqrt{\mu_{2}a_{2}}}\left[(1-e_{1}^{2}) \cos i_{12} {\mathbf e}_{2} \times\ \hat{\mathbf k}_{1} - 5( \mathbf e_{1}\cdot \hat{\mathbf k}_{2}) \mathbf e_{2}\times \mathbf e_{1}\right. \nonumber\\
&& \left. -[\frac{1}{2}-3e_{1}^{2}+\frac{25}{2}(\mathbf e_{1}\cdot \hat{\mathbf k}_{2})^{2}-\frac{5}{2}(1-e_{1}^{2})\cos^{2} i_{12}]\hat{\mathbf k}_{2} \times\mathbf e_{2}\right]\nonumber\\
&& -\frac{25}{16}\epsilon_{\rm oct}\gamma{\sqrt{1-e_{2}^{2}}}\frac{\beta_{1}\sqrt{\mu_{1}a_{1}}}{\beta_{2}\sqrt{\mu_{2}a_{2}}}\Bigg\{2(1-e_{1}^{2})\left[(\mathbf e_{1}\cdot \hat{\mathbf k}_{2})(\hat{\mathbf k}_{1}\cdot \mathbf e_{2})\hat{\mathbf e}_{2}\right.\nonumber\\
&& \left. +\frac{1-e_{2}^{2}}{e_{2}}[\frac{8}{5}\mathbf e_{1}^{2}-\frac{1}{5} - 7(\mathbf e_{1}\cdot \hat{\mathbf k}_{2})
^{2} + (1-e_{1}^{2})\cos^{2} i_{12}]\hat{\mathbf k}_{2}\right]\times \mathbf e_{1}\nonumber\\
&&-\left[2(\frac{1}{5}-\frac{8}{5}\mathbf e_{1}^{2})(\mathbf e_{1}\cdot \hat{\mathbf e}_{2})\mathbf e_{2} + 14(1-e_{1}^{2})(\mathbf e_{1}\cdot \hat{\mathbf k}_{2})(\hat{\mathbf k}_{1}\cdot \hat{\mathbf e}_{2})(\hat{\mathbf k}_{1}\cdot \hat{\mathbf k}_{2})\mathbf e_{2}\right.\nonumber\\
&&\left.+7(\mathbf e_{1}\cdot \hat{\mathbf e}_{2})[\frac{8}{5}\mathbf
 e_{1}^{2}-\frac{1}{5} - 7(\mathbf e_{1}\cdot \hat{\mathbf k}_{2})^{2} +
 (1-e_{1}^{2})\cos^{2} i_{12}]\mathbf e_{2}\right]\times\hat{\mathbf k}_{2})\Bigg\}
\end{eqnarray}
\begin{eqnarray} 
\dot{\mathbf L}_{i}=\alpha_{1i}\cos\theta_{i} \
\hat{\mathbf s}_{i}\times\hat{\mathbf k}_{1}
+\alpha_{2i}\cos\epsilon_{i} \hat{\mathbf s}_{i}\times\hat{\mathbf k}_{2}.
\end{eqnarray}
where
\begin{eqnarray} 
\epsilon_{\rm oct}&=&\frac{m_{0}-m_{1}}{m_{0}+m_{1}}
\frac{a_{1}}{a_{2}}\frac{e_{2}}{1-e_{2}^{2}},\\
\alpha_{1i}&=&\frac{3Gm_{0}m_{1}J_{2i}R_{i}^{2}}
{2a_{1}^{3}(1-e_{1}^{2})^{3/2}},\\
\alpha_{2i}&=& \frac{3Gm_{2}m_{i}J_{2i}R_{i}^{2}}
{2a_{2}^{3}(1-e_{2}^{2})^{3/2}},\\
 \gamma &=& \frac{3Gm_{2}\beta_{1}a_{1}^{2}}{4a_{2}^{3}(1-e_{2}^{2})^{3/2}}. 
\end{eqnarray}
In the above expressions, the parameter $ \epsilon_{\rm oct}$ quantifies
the importance of the octupole term relative to the quadrupole
term.

The magnetic braking as a spin-down process of the central star is
modelled as
\begin{equation} 
\dot {\mathbf L}_{0,\rm mb}
= -\alpha_{\rm mb}C_{0}\omega_{0}^{3}\hat{\mathbf s}_{0},
\end{equation}
where the spin-down rate $\alpha_{\rm mb}$ is set to be $1.66 \times
10^{-13}$ yr according to \citet{Barker2009}. Incidentally the same
magnetic braking effect was incorporated in \citet{Xue2014} although it
was not noted explicitly.

The correction due to general relativity induces the precession of the
pericenter:
\begin{equation} 
\dot {\mathbf e}_{1,\rm GR}=\frac{3\mu_{1}n_{1}}{c^{2}a_{1}(1-e_{1}^{2})}
 \hat{\mathbf k}_{1}\times \mathbf e_{1},
\end{equation}
where $c$ is light speed, and $n_{1}$ is mean motion of the inner orbit.

For the tidal effect, we adopt the equilibrium tidal model with constant
delay time $\Delta{t_{i}}$ \citep{Mignard1979}. Similarly, the averaged
equations are
\begin{equation} 
\dot{\mathbf G}_{2, \rm tide}=0, \qquad 
\dot{\mathbf G}_{1, \rm tide}=-\dot{\mathbf
 L}_{0}-\dot{\mathbf L}_{1},
\end{equation}
\begin{eqnarray} 
\dot{ \mathbf e}_{1, \rm tide}
&=&\sum_{i}\frac{15}{2}k_{2_{i}}n_{1}
\left(\frac{m_{(1-i)}}{m_{i}}\right)
\left(\frac{R_{i}}{a_{1}}\right)^{5}
f_{4}(e_{1})\hat{\mathbf k}_{1}\times \mathbf e_{1}\nonumber\\
&& -\sum_{i}\frac{K_{i}}{\beta_{1}a_{1}^{2}}
\left[f_{4}(e_{1})\frac{\omega_{i}}{2n_{1}}( \mathbf e_{1}\cdot \hat{\mathbf s}_{i})\hat{\mathbf k}_{1}-\left(\frac{11}{2}f_{4}(e_{1})\cos\theta_{i}\frac{\omega_{i}}{n_{1}}-9f_{5}(e_{1})\right) \mathbf e_{1}\right],\\
\dot{\mathbf L}_{i, \rm tide}
&=&K_{i}n_{1}\left[ f_{4}(e_{1})\sqrt{1-e_{1}^{2}}\frac{\omega_{i}}{2n_{1}}(\hat{\mathbf s}_{i}-\cos\theta_{i}\hat{\mathbf k}_{1})\nonumber \right. \\
 &&\left. -f_{1}(e_{1})\frac{\omega_{i}}{n_{1}}\hat{\mathbf s}_{i}+f_{2}(e_{1})\hat{\mathbf k}_{1}+\frac{(\mathbf e_{1}\cdot \hat{\mathbf s}_{i})(6+e_{1}^{2})}{4(1-e_{1}^{2})^{9/2}}\frac{\omega_{i}}{n_{1}}\mathbf e_{1} \right].
\end{eqnarray}
where
\begin{equation} 
K_{i}=\Delta{t_{i}}\frac{3k_{2_{i}}Gm_{(1-i)}^{2}R_{i}^{5}}{a_{1}^{6}},
\end{equation}

\begin{eqnarray}
 f_{1}(e) &=& \frac{1+3e^{2}+3e^{4}/8}{(1-e^{2})^{9/2}},\\
 f_{2}(e) &=& \frac{1+15e^{2}/2+45e^{4}/8+5e^{6}/16}{(1-e^{2})^{6}},\\
 f_{3}(e) &=& \frac{1+31e^{2}/2+255e^{4}/8+185e^{6}/16+25e^{8}/64}{(1-e^{2})^{15/2}},\\
 f_{4}(e) &=& \frac{1+3e^{2}/2+e^{4}/8}{(1-e^{2})^{5}},\\
 f_{5}(e) &=& \frac{1+15e^{2}/4+15e^{4}/8+5e^{6}/64}{(1-e^{2})^{13/2}}. 
\end{eqnarray}

\section{Short-range force effects: Precession rate on $\hat{\mathbf e}_{1}$
\label{sec:srf-eqs}}

The three main short-range forces (GR, planetary tide and rotational
distortion) modify $\hat{\mathbf e}_{1}$, and induce an additional
precession of $\hat{\mathbf e}_{1}$ around $\hat{\mathbf k}_{1}$:
\begin{equation} 
\dot {\mathbf e}_{1} = \omega_{\rm pre} \hat{\mathbf k}_{1}
\times \mathbf e_{1}.
\end{equation}

The precession rate, $\omega_{\rm pre}$, for the three main short-range
forces can be read off from the evolution equations in
\ref{sec:correia-eqs} in a straightfoward manner as
\begin{eqnarray}
\label{eq:omegaGR}
\omega_{\rm GR} &=& \frac{3\mu_{1}n_{1}}{c^{2}a_{1}(1-e_{1}^{2})}
\propto \frac{1}{a_{1}^{5/2}(1-e_{1}^2)}, \\
\label{eq:omegaPT}
\omega_{\rm PT} &=&
 \frac{15}{2}k_{2_{1}}n_{1} \left(\frac{m_{0}}{m_{1}}\right)
\left(\frac{R_{1}}{a_{1}}\right)^{5}
f_{4}(e_{1})\propto \frac{1}{a_{1}^{13/2}(1-e_{1}^2)^{5}}, \\
\label{eq:omegaPRD}
\omega_{\rm PRD} &=& \frac{\alpha_{11}}{||\mathbf
 G_{1}||}\frac{1}{2}(1-5\cos^{2}\theta_{1})
\propto \frac{1}{a_{1}^{7/2}(1-e_{1}^2)^{2}}, 
\end{eqnarray}
where $\omega_{\rm GR}, \omega_{\rm PT},$ and $\omega_{\rm PRD}$ are
the precession rate induced by GR, planetary tide, and planetary
rotational distortion, respectively. 

Note that the above expressions are consistent with those of
\citet{Liu2015} if the tidal Love number is set to be twice of the
deformation love number $k_{2_{i}}$, and the spin and orbit of the inner
planet are aligned ($\theta_{1}=0^{\circ}$).


\section{Effect of the spin rotation period of the inner planet
\label{sec:w10h}}

\begin{figure}[h]
\begin{center}
\includegraphics[width=13cm]{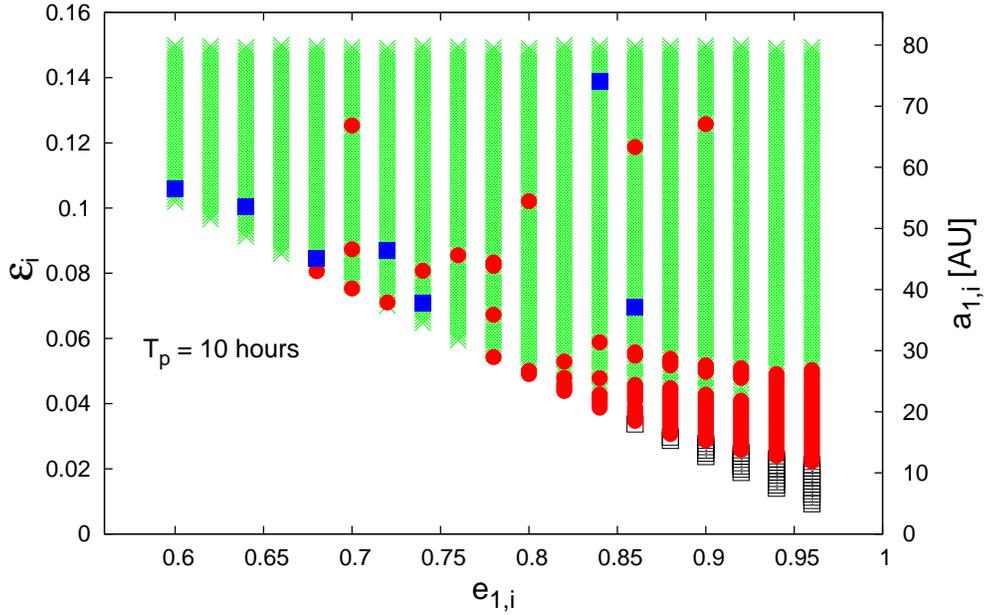} 
\caption{Same as Figure \ref{fig:a2500} but with $T_{p}=10$ hours as the
 planetary spin rotation period.}
\label{fig:w10}
\end{center}
\end{figure}

Throughout the present analysis, we have adopted 10 days as the spin
rotation period of the inner planet. If one considers Jupiter as a typical
planet, 10 hours, instead of 10 days, may be more relevant.  Therefore
we repeat our fiducial run using the 10 hour period while keeping all the
other parameters unchanged.  Figure \ref{fig:w10} shows the result,
which is basically identical with Figure \ref{fig:a2500}. Just for
more quantitative comparison, we show the branching ratios of the final
outcomes; PHJ 8.5\%, RHJ 0.4\%, NM 2.1\%, and TD 89.0\%.  Thus we
conclude that the final result is very insensitive to the choice of the
planetary spin period in this range.

\end{document}